\documentclass[preprint,flushrt]{aastex1} 
\usepackage[varg]{txfonts}
\usepackage{color}
\usepackage{amssymb}
\usepackage{graphicx, natbib}
\bibpunct{(}{)}{;}{a}{}{,}
\newcommand{\kms}{~km~s$^{-1}$}
\newcommand{\WB}{_{\mathrm{WB}}}
\shorttitle{Flow vector, Mach number, and abundance of the Warm Breeze}
\shortauthors{Kubiak et al.}

\begin{document}

\title{Interstellar neutral helium in the heliosphere from IBEX observations. IV. Flow vector, Mach number, and abundance of the Warm Breeze}
\author{Marzena A. Kubiak\altaffilmark{1}, P.~Swaczyna\altaffilmark{1}, M.~Bzowski\altaffilmark{1}, J.~M.~Sok{\'o}{\l}\altaffilmark{1}, S.~A.~Fuselier\altaffilmark{2,3}, A.~Galli\altaffilmark{4}, D.~Heirtzler\altaffilmark{5}, H.~Kucharek\altaffilmark{5}, T.~W.~Leonard\altaffilmark{5}, D.~J.~McComas\altaffilmark{2,3}, E.~M{\"o}bius\altaffilmark{5}, J.~Park\altaffilmark{5}, N.~A.~Schwadron\altaffilmark{5}, P.~Wurz\altaffilmark{4}}
\email{mkubiak@cbk.waw.pl}

\altaffiltext{1}{Space Research Centre of the Polish Academy of Sciences (CBK PAN), 00-716 Warsaw, Poland}
\altaffiltext{2}{Southwest Research Institute, San Antonio, TX, USA}
\altaffiltext{3}{University of Texas at San Antonio, San Antonio, TX, USA}
\altaffiltext{4}{Physikalisches Institut, Universit{\"a}t Bern, Bern, Switzerland}
\altaffiltext{5}{Space Science Center and Department of Physics, University of New Hampshire, Durham, NH, USA}

\date{Received  / Accepted }

\begin{abstract}
With the velocity vector and temperature of the pristine interstellar neutral (ISN) He recently obtained with high precision from a coordinated analysis summarized by \citet{mccomas_etal:15b}, we analyzed the IBEX observations of neutral He left out from this analysis. These observations were collected during the ISN observation seasons 2010---2014 and cover the region in the Earth's orbit where the Warm Breeze persists. We used the same simulation model and a very similar parameter fitting method to that used for the analysis of ISN He. We approximated the parent population of the Warm Breeze in front of the heliosphere with a homogeneous Maxwell-Boltzmann distribution function and found a temperature of $\sim 9\,500$~K, an inflow speed of 11.3~\kms, and an inflow longitude and latitude in the J2000 ecliptic coordinates $251.6\degr$, $12.0\degr$. The abundance of the Warm Breeze relative to the interstellar neutral He is 5.7\% and the Mach number is 1.97. The newly found inflow direction of the Warm Breeze, the inflow directions of ISN H and ISN He, and the direction to the center of IBEX Ribbon are almost perfectly co-planar, and this plane coincides within relatively narrow statistical uncertainties with the plane fitted only to the inflow directions of ISN He, ISN H, and the Warm Breeze. This co-planarity lends support to the hypothesis that the Warm Breeze is the secondary population of ISN He and that the center of the Ribbon coincides with the direction of the local interstellar magnetic field. The common plane for the direction of inflow of ISN gas, ISN H, the Warm Breeze, and the local interstellar magnetic field 
is given by the normal direction: ecliptic longitude $349.7\degr \pm 0.6\degr $ and latitude $35.7\degr \pm 0.6$ in the J2000 coordinates, with the correlation coefficient of 0.85. 
\end{abstract}


\section{Introduction}

Observations of neutral gas by the Interstellar Boundary Explorer \citep[IBEX,][]{mccomas_etal:09a} provide important insight into the physical state and processes operating in the interstellar matter in front of the heliosphere. So far, IBEX has observed interstellar hydrogen, helium, oxygen, neon, and deuterium \citep{mobius_etal:09b, bochsler_etal:12a, saul_etal:12a, park_etal:14a, park_etal:15a, rodriguez_etal:13a}. 

Based on IBEX observations, \citet{kubiak_etal:14a} discovered a previously unknown population of neutral helium in the heliosphere, which they dubbed the Warm Breeze (WB). This population is most visible in the portion of Earth's orbit just before the region where interstellar neutral helium (ISN He) is observed. Based on analysis of data collected by IBEX over a single season of ISN gas observations, they reported that the source of WB can be reasonably approximated by a homogeneous Maxwell-Boltzmann population of neutral He gas in a region $\sim 150$~AU in front of the heliosphere and determined the best fitting temperature $T\WB$, inflow direction (ecliptic longitude $\lambda\WB$ and latitude $\beta\WB$), speed $v\WB$, and abundance $\xi\WB$ relative to the primary ISN He within relatively broad uncertainties: $\lambda\WB = 240\degr \pm 10\degr$, $\beta\WB = 11\degr^{+7\degr}_{-3\degr} $, $v\WB = 11 \pm 4$\kms, $T\WB = 15\,000^{+6000}_{-8000}$~K, and $\xi\WB = 0.07 \pm 0.03$. They also pointed out that the fit quality obtained was not satisfactory as indicated by a large reduced chi square value of $\sim 4$ and reported that the signal for some IBEX spin angles depended heavily on a hypothetic threshold in the sensitivity of the IBEX-Lo instrument to low-energy neutral He atoms. 

One of the most important conclusions suggested by \citet{kubiak_etal:14a} is that the Warm Breeze may be the secondary population of the ISN He gas. The secondary population of heliospheric neutrals is created in the outer heliosheath, where the originally unperturbed flow of interstellar plasma is deflected to flow past the heliopause, which is an impenetrable obstacle for interstellar plasma ions. On the other hand, the neutral component of ISN gas is collisionless on spatial scales comparable to the size of the heliosphere and it is not subject to electromagnetic forces governing the plasma, so it continues its bulk motion almost without modifications. This causes decoupling of the ionized and neutral component flows. The ionized component is compressed and heated while flowing past the heliosphere, which enhances charge exchange collisions between the perturbed plasma and pristine neutral flows. As a result, some ions that belonged to interstellar plasma become neutralized, and some atoms from the neutral component become ionized and picked up by the plasma flow. Since resonant charge exchange reactions operate practically without momentum exchange between the collision partners, the new population of neutralized interstellar ions inherits the local parameters of the ambient plasma, which are different from the parameters of the unperturbed interstellar gas, and continue flowing away from their birth location, decoupled from the parent plasma. Some of those atoms enter the heliosphere, where they are subject to gravitational acceleration and ionization. Since the ionization losses in He atoms are relatively small \citep{bzowski_etal:13a}, an appreciable fraction of the secondary He atoms penetrate into the Earth's orbit, where they are measured by IBEX. 

This mechanism of creation of the heliospheric secondary neutral population has been anticipated theoretically for quite a while \citep{baranov_etal:81a, baranov_malama:93}. The secondary population of ISN H has been believed to exist based on many observations carried out using various techniques (see \citet{katushkina_etal:15b} for a recent review), but -- to our knowledge -- has not previously been unambiguously resolved from the primary population. The secondary He population had been believed to be of negligible abundance \citep{mueller_zank:04a} because the reaction assumed to be responsible for its creation: He$^+$ + H $\rightarrow $ He + H$^+$ has a very low cross section \citep{barnett_etal:90}, in contrast to similar interactions with oxygen ions. However, \citet{bzowski_etal:12a} pointed out that the cross section for the charge exchange reaction between neutral He atoms and He$^+$ ions is comparable to the large cross section for the charge exchange between H atoms and protons, and because of the relatively high abundance of He$^+$ ions in the interstellar gas near the heliosphere \citep{frisch_slavin:03} appreciable amounts of the secondary He atoms should be produced in the outer heliosheath.

If the Warm Breeze is indeed the secondary population of ISN He, then it provides information about the physical state of interstellar matter in the outer heliosheath because it can be clearly separated from the primary population. Based on analysis of the secondary population one can infer the temperature, flow speed, and the deflection angle of the flow direction of the secondary component due to the deformation of the heliosphere from axial symmetry by the action of the interstellar magnetic field (ISMF). It was suggested by \citet{lallement_etal:05a} and found from different heliospheric models \citep[e.g.,][]{izmodenov_etal:05a, pogorelov_etal:08a} that the secondary population of heliospheric neutrals should have a flow velocity vector in the plane formed by the inflow direction of the unperturbed interstellar matter and the unperturbed vector of ISMF. Thus, if we are able to determine the inflow direction of the secondary component of the ISN gas, then, with the inflow vector of the unperturbed interstellar gas available \citep{witte:04, bzowski_etal:14a, bzowski_etal:15a, leonard_etal:15a, mccomas_etal:15a, mccomas_etal:15b, schwadron_etal:15a, wood_etal:15a}, we can also determine the plane in which the ISMF vector is expected to be. Constraining the ISMF vector reduces the number of unknown parameters which hamper heliospheric studies using large simulation codes, like the Moscow Monte Carlo model \citep{izmodenov_alexashov:15a}, the Huntsville model \citep{pogorelov_etal:09c}, or the University Michigan/Boston University model \citep{opher_etal:06a}.

If, on the other hand, the Warm Breeze is not the secondary population of ISN He, then an even more compelling question appears: what is its nature and origin. Answering these questions is only possible by a more thorough analysis of the available data, which is the topic of this article. We analyze the Warm Breeze observations carried out by IBEX out during the ISN observation seasons from 2010 through 2014 and derive its temperature, abundance, and inflow velocity vector. Based on those parameters and their relation to other heliospheric observables, we discuss possible sources for this population.

\section{Observations and data selection}

\begin{figure*}
\centering
\includegraphics[width=1.0 \textwidth, height=0.8\textheight]{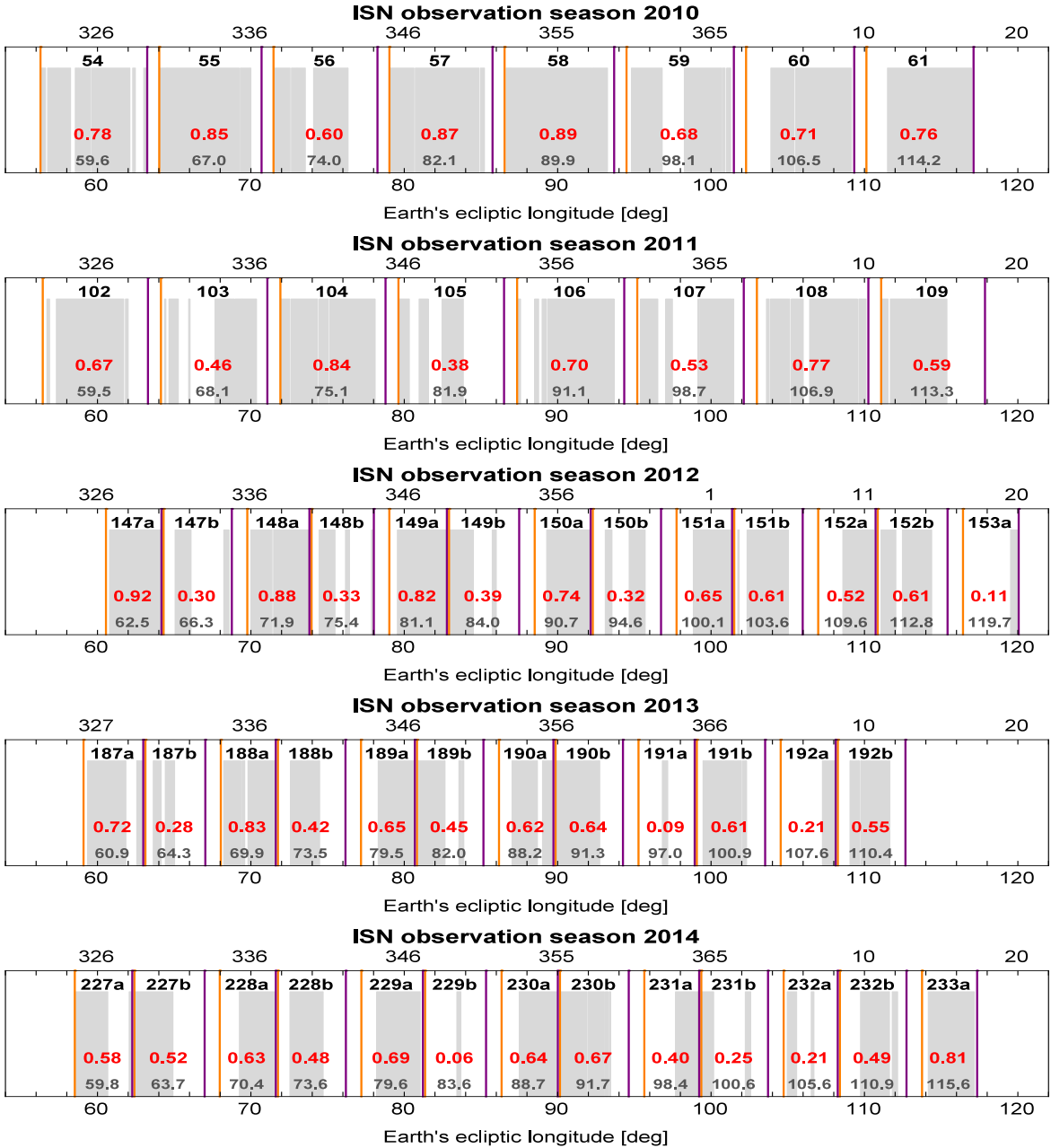}
\caption{Distribution of good time intervals during the IBEX-Lo observations of the Warm Breeze adopted for the analysis. The format of the figure is similar to that of  Figure 1 in \citet{bzowski_etal:15a}. The gray regions mark the individual good times intervals. The orange bars mark the beginning and the purple bars the end of the High-Altitude Science Operations (HASO) intervals, when the IBEX measurements were actually carried out. The thick black labels mark individual orbits (or orbital arcs, in the 2012---2014 seasons). The red numbers (in the middle row) mark the fraction of HASO intervals occupied by the good time intervals for a given orbit. In the lower row, the approximate longitudes of the spacecraft during a given orbit are marked (actually, it is the Earth longitude averaged over the ISN good times for an orbit). They can be used to identify an approximate correspondence between orbits from different seasons. The lower horizontal axes are scaled in the Earth's ecliptic longitude and the upper horizontal axes are scaled in days of the calendar year (note that new year begins during each individual Warm Breeze observation season.)}
\label{fig:goodTimes}
\end{figure*}

The strategy of ISN observations by IBEX has been presented by \citet{mobius_etal:09a, mobius_etal:12a}, and \citet{mobius_etal:15a}. The details the most relevant for the present analysis are discussed in Section~2 in \citet{swaczyna_etal:15a}; here we only point out the most important aspects. 

IBEX is a sun-pointing spinning spacecraft \citep{mccomas_etal:09a} on a highly elongated elliptical orbit around the Earth \citep{mccomas_etal:11a}, and the IBEX-Lo time-of-flight mass spectrometer, used for the ISN atom observations \citep{fuselier_etal:09b}, scans a great circle on the sky perpendicular to the spin axis. The instrument has eight logarithmically spaced energy channels of wide acceptance ($\Delta E/E \simeq 0.7$), which are sequentially switched during operation. ISN atoms are observed over a few month interval around the beginning of each calendar year, when the spacecraft together with the Earth move toward the ISN flow, thus increasing the relative speed (and energy) of the atoms and consequently their flux and the efficiency of their detection. As a result of this observation geometry, the relative energy of neutral He atoms varies during the observation season and also as a function of the spacecraft spin angle. Neutral atoms enter the instrument through a collimator and hit a specially prepared carbon conversion surface, which retains a very thin layer of absorbed material which is mostly water. Some of the species measured by IBEX-Lo can be identified directly because on hitting the conversion surface they form negative ions, which are then extracted by an electric field, accelerated, and analyzed by the electrostatic analyzer. Noble gases like He and Ne, however, do not form stable negative ions and therefore can be observed only indirectly. When these neutrals hit the conversion surface, they sputter a cloud of negative C, O, and H ions \citet{wurz_etal:08a}, which are collected by the electrostatic analyzer and registered by the time-of-flight spectrometer. Species identification is carried out on ground, by analysis of the proportions between the time-of-flight signals of the sputtered C, H, and O atoms. Details of the measurement process and data flow are presented by \citet{mobius_etal:15a, mobius_etal:15b}, and details of the species identification by \citet{park_etal:14a, park_etal:15a}. The sputtering products have energies lower than the energy of the incident atom. The energy spectrum of the sputtering products is relatively flat between 0 eV and a drop off at an energy a little lower than the energy of the incoming neutral atom. In addition, there is a finite minimum energy for an incoming atom to sputter, which we refer to as the energy threshold for sputtering. Therefore, the sputtering products are registered in all energy channels between the lowest channel and the channel with the energy acceptance corresponding to the energy of the incoming atom. The most abundant species among the sputtering products of He is hydrogen. H ions sputtered by He are observed mostly in IBEX-Lo energy channels 1 through 3. Without further analysis of the proportions of the H$^-$ signal to the signal from the simultaneously registered C$^-$ and O$^-$ ions, the H$^-$ ions sputtered by incident He atoms are indistinguishable from those produced by the incoming H atoms.

The Warm Breeze is most visible from mid-November to the end of January each year. In this portion of the Earth's orbit, the Warm Breeze signal observed by IBEX is relatively little affected by the primary population of ISN He \citep{kubiak_etal:14a, sokol_etal:15a}, and its signal is only slightly modified by the magnetospheric foreground, if data selection is carried out carefully \citep{galli_etal:14a, galli_etal:15a}. In the present analysis, we used observations from this portion of the Earth's orbit. To maintain as much year-to-year repeatability of the observation conditions as possible, we adopted a common criterion for the orbit selection: we chose the IBEX orbits with the spin axis pointing within the range of ecliptic longitudes $(235\degr, 295\degr)$, i.e., from mid-November of the year preceding the given season year to the end of January. This choice effectively included all orbits where the Warm Breeze signal is clearly visible and by design it ends at the beginning of the range chosen by \citet{bzowski_etal:15a} for the analysis of the primary ISN He. The data used were collected during the IBEX ISN gas observation campaigns from 2010---2014, because the commissioning of the spacecraft during the 2009 IBEX ISN season was completed too late to observe the Breeze. We use the Histogram-Binned data product, corrected where needed for the instrument throughput reduction \citep{mobius_etal:15a, swaczyna_etal:15a}, and take the Golden Triples events, i.e., we only take those events with three time-of-flight measurements that are almost certainly due to H$^-$ ions. The data are binned into 60 equal-width bins covering the full $360\degr$ range of IBEX spin angles. A connection between the IBEX spin angles and the absolute directions in the sky for individual orbits is presented in Figure 2 in \citet{sokol_etal:15a} (see also the transformation matrix given in Equation (22) in \citet{sokol_etal:15b}, and the spin axis orientation in \citet{swaczyna_etal:15a}.)
 
Selection of data from individual orbits was carried out using the same criteria as for the ISN He observations reported by \citet{bzowski_etal:15a}, \citet{leonard_etal:15a}, \citet{mccomas_etal:15a}, \citet{mobius_etal:15a}, \citet{schwadron_etal:15a}, and \citet{swaczyna_etal:15a}. A summary of this coordinated analysis is given by \citet{mccomas_etal:15b}. In brief, these criteria rejected the data intervals with synchronization issues, with known magnetospheric contamination, and with excessive signals observed in other IBEX energy channels. Details of good times selection were presented by \citet{fuselier_etal:14a}, \citet{galli_etal:14a} and \citet{galli_etal:15a}, as well as \citet{leonard_etal:15a} and \citet{mobius_etal:15a}. The good times intervals used in our analysis are presented in Figure~\ref{fig:goodTimes}.

\begin{figure*}
\centering
\includegraphics[width=0.49 \textwidth]{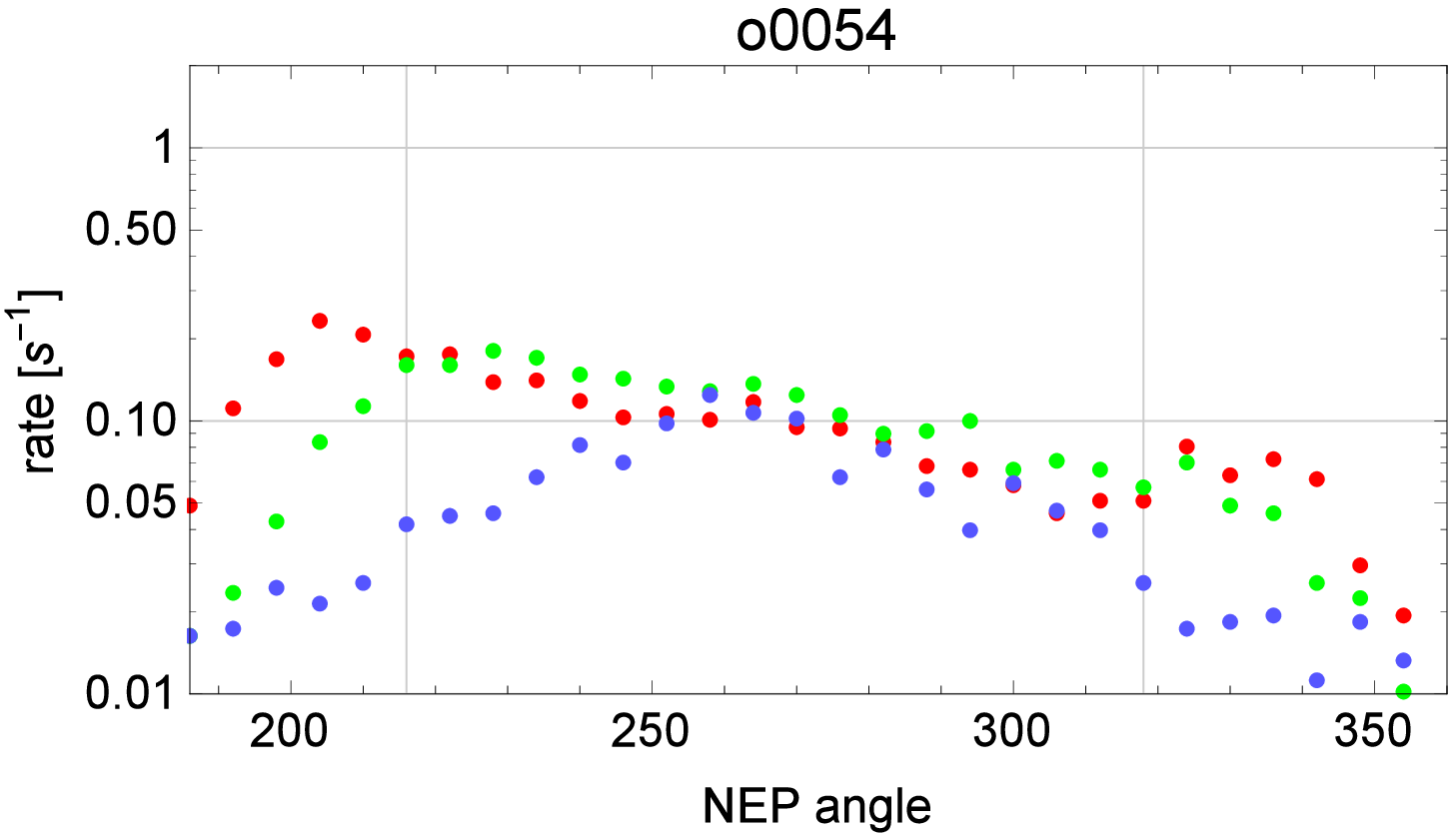}
\includegraphics[width=0.49 \textwidth]{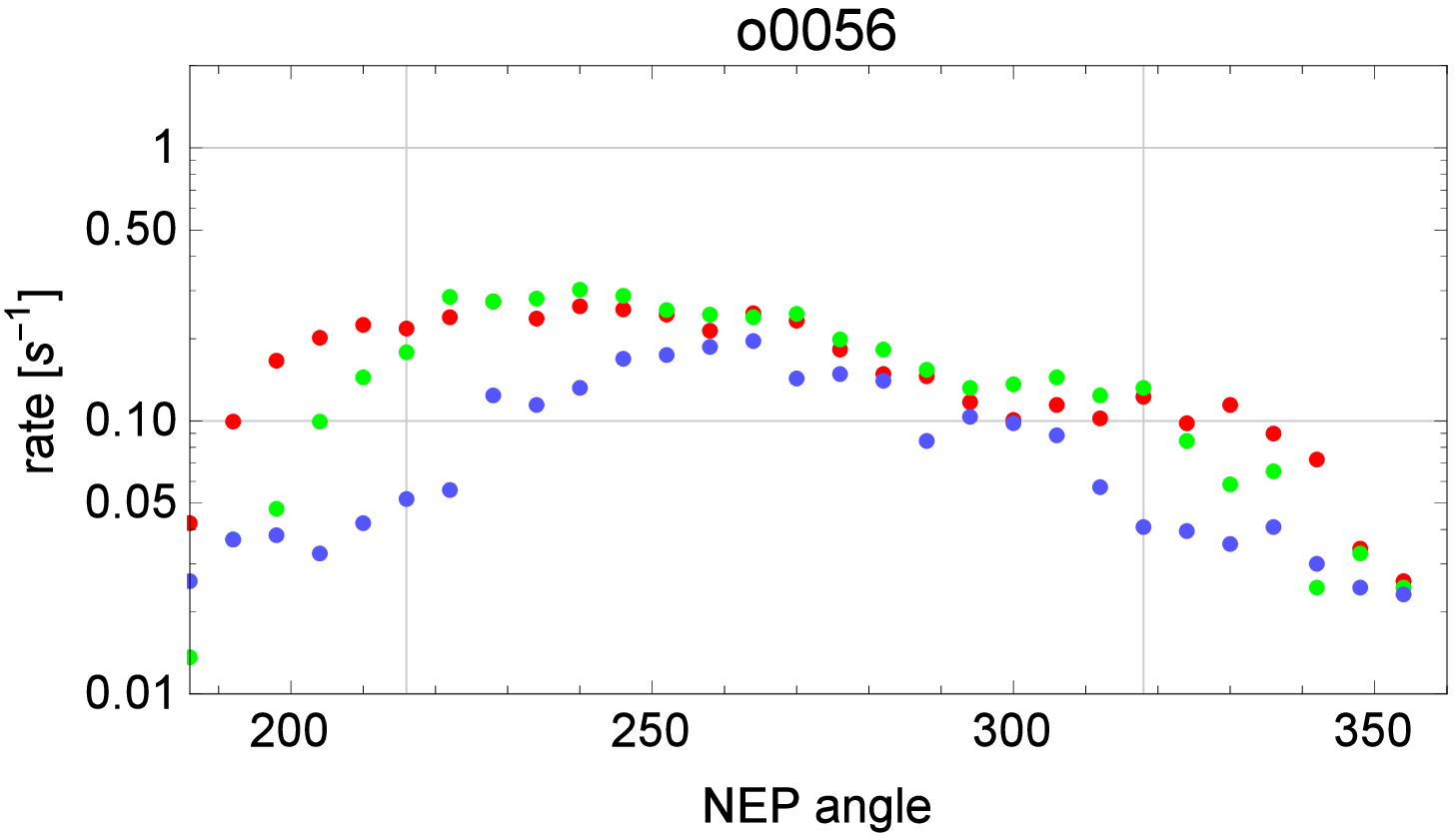}

\includegraphics[width=0.49 \textwidth]{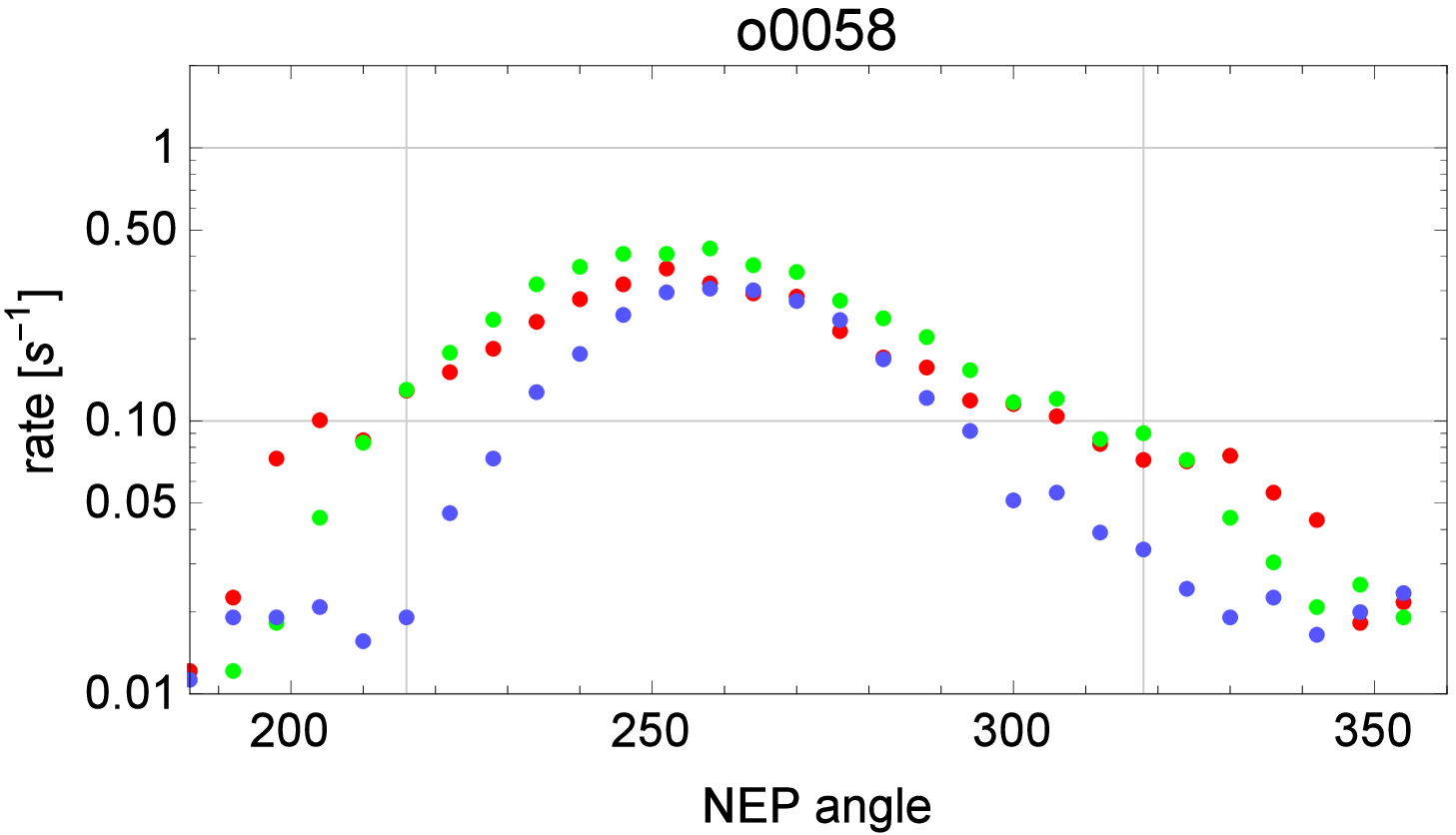}
\includegraphics[width=0.49 \textwidth]{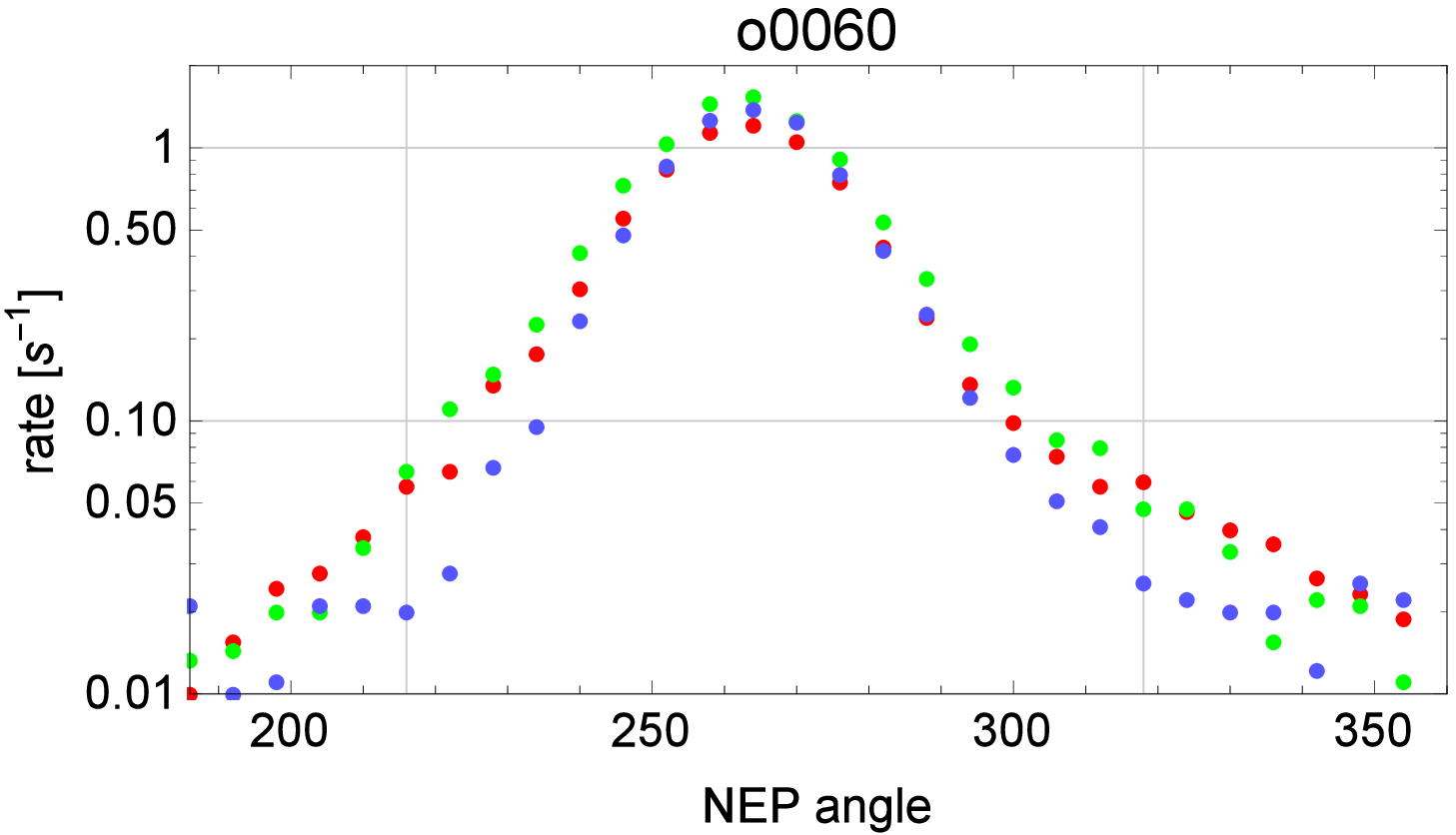}
\caption{Count rates observed in four orbits from the Warm Breeze observation season 2010 as a function of spin angle. Unlike in Figure \ref{fig:residuals}, no background and ISN He subtraction has been done. Each dot represents the good-times averaged count rate for an individual 6-degree pixel. The colors symbolize different IBEX-Lo energy steps: red -- channel 1, green -- channel 2 (used in the fitting), blue -- channel 3. The vertical bars mark the interval of spin angles selected for the baseline fitting. The interval shown is from spin angle $180\degr$ to $354\degr$, i.e., the wider interval used in the complementary fit (see text). The four orbits are chosen as illustrative examples out of the eight orbits from this observation season used in the analysis. Note that the count rate variations with the spin angle for energy channels 1 and 2 are similar to each other while energy channel 3 features clear deviations from the other two channels. The behavior of the data from the other observation seasons is similar. }
\label{fig:data3Channel}
\end{figure*}

The final step in our data selection was choosing the spin angle range. The Warm Breeze signal is visible in energy channels 1, 2, and 3, but the signals for individual energy channels differ from each other (Figure~\ref{fig:data3Channel}). These differences are most likely due to different energy sensitivities of IBEX-Lo in different energy channels. On one hand, the absolute levels of the signal vary from one energy channel to another, and on the other, the fall off in the wings of the signal seems to start at the spin angles that depend on the energy of the incoming atom. In principle, such differences are expected, as shown by modeling by \citet{kubiak_etal:14a} and \citet{sokol_etal:15a} and observed in the measurements by \citet{galli_etal:15a}. The signal falls off for those spin angles where atoms with lower energies enter the instrument, but for each orbit there is an interval of spin angles where the flux does not depend on the adopted energy threshold regardless of the magnitude of the latter within all reasonably expected values.

A potentially important difference between our simulations and the actual measurement process is that while the simulation calculates the flux of neutral He atoms hitting the instrument and its collimation by the IBEX-Lo collimator \citep{sokol_etal:15b}, it does does not emulate any processes related to the conversion of the He atom flux into the count rate of the H$^-$ ions that IBEX-Lo registers other than a sharp cutoff at the lower end of the energy spectrum of the incoming atoms. The He atoms hitting the instrument predominantly have kinetic energies much larger than the boundaries of the three lowest energy channels of IBEX-Lo, and the sputtered H$^-$ ions are expected to have a roughly flat energy  spectrum in the energy range corresponding to the energy ranges of at least IBEX energy channel 1 and 2 (see Figure~1 in \citet{mobius_etal:12a} and \citet{saul_etal:12a}). Therefore it is expected that even though the absolute magnitudes of the count rates measured in channels 1 and 2 may systematically differ, the shapes of the signal as a function of spin angle should be very similar. Departures may suggest that some of the signal is not due to sputtering by He atoms (e.g., a local foreground) or that the spectrum of the sputtering products is not flat, e.g., because of a finite energy threshold for the sputtering. Therefore we chose the spin angle range where the count rates in energy channels 1 and 2 tracked each other well (i.e., we rejected the parts where the signal started to precipitously fall off, as expected for a finite energy threshold, see Figures~8 and 9 in \citet{kubiak_etal:14a}, and Figure~10 in \citet{sokol_etal:15a}). Another consideration was finding a common spin angle range for all orbits included in this study. This was important because on one hand we suspect the data may still contain some remnant foreground, and on the other hand the width of the signal in the spin angle space narrows towards later orbits. Maintaining identical numbers of data points for all orbits guarantees that no orbit is statistically biasing the results because of the number dominance of data points it contributes to the global sample. These considerations resulted in adoption of data from energy channel 2 from the spin angle range from $216\degr$ to $318\degr$, i.e., 18 data points per orbit. The data selected are shown in the upper panes of the panels in Figures \ref{fig:residuals}.

In principle, we could have chosen energy channel 1 instead of 2. However, as shown by \citet{saul_etal:12a}, energy channel 1 in the ISN orbit range contains an appreciable contribution from ISN H, with channel 2 affected much less \citep[see also][]{schwadron_etal:13a, katushkina_etal:15b}. This component is not expected in the orbits from the Warm Breeze range (however, see the {\it Discussion} later in the text), but to infer the abundance of the Warm Breeze we must use the scaling factors between the simulated flux and the measured count rate, obtained for ISN He for each observation season. These factors were found in the fitting of the ISN He parameters by \citet{bzowski_etal:15a}, who used data from energy channel 2. 

To test the robustness of our results, we also used data from an extended and narrowed spin angle ranges: $204\degr$---$330\degr$ and $228\degr$---$306\degr$, respectively, and additionally, the spin angle range covering the entire ram hemisphere: $180\degr$---$354\degr$. In the two extended ranges, we do expect some dependence of the signal on the sputtering threshold and therefore throughout the simulations we adopted a value for this threshold equal to 38~eV, consistent with the results of the analysis by \citet{galli_etal:15a}.

\section{Analysis}
In this section we present the analysis process. We start with a presentation of the physical model for the Warm Breeze phenomenon we adopted, the preparation of the data selected in the previous section for parameter fitting, and the aspects of the measurement process that affect the data measurement uncertainty and correlations between individual data points. Then we discuss the method of parameter fitting and assessing their uncertainties, with the correlations between the parameters presented. We also discuss the residuals and their implications. Finally, we present additional tests of the robustness of the results and derive the uncertainties and correlations of the Warm Breeze parameters.

\subsection{The physical model, uncertainty system, and data preparation}
The analysis was carried out using the method, data uncertainties, and data correlation system presented in detail by \citet{swaczyna_etal:15a} These aspects of the analysis are similar to those in the determination of the inflow parameters of the primary ISN He population by \citet{bzowski_etal:15a}. All of the data used had been corrected for the throughput reduction in the instrument interface, following the scheme presented in detail by \citet{swaczyna_etal:15a}, except for the data from the 2013 and 2014 seasons, for which the correction is not needed owing to an on-board software change. The magnitudes of the correction and their uncertainties were calculated based on the actually measured data\footnote{For some ISN orbits in the 2011 and 2012 seasons, the precise value of the correction could not be calculated due to a special observation mode of the instrument, as explained by \citet{mobius_etal:15b}, and an average value was used instead, but for the data used here in the Warm Breeze parameter fitting, no such approximate measures had been needed.}. 

The uncertainty system used in the calculation of the data covariance matrix includes the statistical uncertainty of the Poisson counting process (uncorrelated between the data points), the background (assumed to be constant for all data points and thus correlating them), the uncertainty of the spin axis (which correlates points from individual orbits), the uncertainty of the IBEX-Lo boresight orientation with respect to the spin axis (identical for all orbits, it correlates all data points), and the uncertainty of the throughput correction. The closing element of the uncertainty system is the uncertainty of the primary ISN He model, which was adopted as obtained from the analysis of ISN He by \citet{bzowski_etal:15a}. Before fitting, we subtracted the simulated signal from the ISN He primary population from the data. Also subtracted was the constant level of the ubiquitous background, adopted after \citet{galli_etal:14a} at $(8.9 \pm 1.0) \cdot 10^{-3}$~cts~s$^{-1}$ for the observation seasons 2010---2012 and to $(4.2 \pm 0.5)\cdot 10^{-3}$ for the 2013 and 2014 seasons, i.e., after the PAC voltage reduction \citep{mobius_etal:15b}, identically as was done by \citet{bzowski_etal:15a}. The ISN He signal was calculated precisely for the actual observation conditions and using the yearly scaling factors that were fitted by \citet{bzowski_etal:15a} together with the ISN He inflow parameters. This was needed to exactly reproduce the observed count rates.   

The ISN He signal subtracted from the data and the signal from the Warm Breeze were simulated using the latest version of the Warsaw Test Particle Model, presented in detail by \citet{sokol_etal:15b} with the time-dependent photoionization rate from \citet{sokol_bzowski:14a}. The physical model of the neutral He gas observed by IBEX is a superposition of a neutral He flux at the Earth's orbit originating from two Maxwell-Boltzmann populations of neutral He atoms in front of the heliosphere: the primary ISN He population, with the temperature and inflow velocity vector in the source region as reported by \citet{bzowski_etal:15a} for the fit to the ISN He data from all seasons ($\lambda_{\mathrm{ISN}} = 255.75\degr$, $\beta_{\mathrm{ISN}} = 5.16\degr$, $v_{\mathrm{ISN}} = 25.76$~\kms, $T_{\mathrm{ISN}} = 7440$~K), and another Maxwell-Boltzmann population, corresponding to the Warm Breeze, with the parameters being sought. These latter parameters include a temperature $T\WB$, a velocity vector ($v\WB, \lambda\WB, \beta\WB$), and an abundance $\xi\WB$ relative to the primary ISN He population. In the actual fitting, the temperature was replaced with the mean square of the thermal velocity, defined as $v_{\mathrm{T,WB}} = \sqrt{3 k_B T\WB/m_{\mathrm{He}}}$. The parameters of both populations were assumed to be homogeneous in the source region in front of the heliosphere, and the distance to the source (i.e., the distance of tracking the test atoms in the model) was set to 150~AU from the Sun. Discussion of the reasons for this choice of the source distance and of its very small influence on the results for the ISN He gas can be found in \citet{mccomas_etal:15b} and \citet{sokol_etal:15b}. The relatively small tracking distance for the Warm Breeze population, which places the source region just beyond the heliopause, is especially reasonable for the hypothesis that the Warm Breeze is the secondary component of ISN gas that would be formed at roughly such a distance, but the parameter values we have obtained are very little sensitive to this choice. We assumed that the velocity vector and temperature of the Warm Breeze in the source region do not change with time, but -- because some measurement aspects varied between the observation seasons, as explained in detail by \citet{bzowski_etal:15a} -- we allowed the abundance parameter to vary from year to year, i.e., we adopted five free abundance parameters, used for the five yearly data subsets. Hence the total number of fit parameters was equal to 9. The abundance parameters were fitted analytically, as described by \citet{sokol_etal:15b}, and were not part of the parameter grid discussed below. 

\subsection{Parameter fitting}

The Warm Breeze parameter fitting was carried out in a two-step process. In the first step, we used a simplified fitting to approximately determine the parameter correlation line for the solution. This was done using a method similar to the method employed by \citet{kubiak_etal:14a} in the Warm Breeze discovery paper. In this method, a simplified uncertainty system was used, including only the statistical uncertainty of the signal and the uncertainty of the background. The fits were carried out to the data with the background and the model of the ISN He population subtracted. They were performed for two spin angle ranges: $222\degr$---$312\degr$ and $186\degr$---$354\degr$.

Subsequently, with the approximate parameter correlation lines established and an approximate best-fit solution found, we defined a regular grid of parameters in the four-dimensional (4D) parameter space to carry out the calculations needed to obtain the data correlation matrix, as described by \citet{swaczyna_etal:15a}. Based on the simplified fitting, we selected the range of ecliptic longitudes between $237\degr$ and $259\degr$ with a step $\Delta\lambda=2\degr$. For the remaining parameters, we decided to adopt the following steps of the grid: $\Delta\beta=0.5\degr$, $\Delta v_{\mathrm{T}}=0.5$~\kms and $\Delta v=0.5$~\kms, and for each ecliptic longitude of the original grid we selected a point $(\beta_0, v_{\mathrm{T}0}, v_0)$ that was nearest to the minimum $\chi^2$ for this longitude and had coordinates being integer multiples of the planned step of the grid. The grid nodes were constructed so that for each longitude, we found points $(\beta_0 + n_{\beta}\, \Delta \beta, v_{\mathrm{T}0} + n_{v_{\mathrm{T}}}\, \Delta v_{\mathrm{T}}, v_0 + n_v \, \Delta n_v)$ such that for integer values of $n_{\beta}$, $n_{v_{\mathrm{T}}}$, $n_v$ the condition $\sqrt{n_{\beta}^2 + n_{v_{\mathrm{T}}}^2 + n_v^2} < 4.5$ was fulfilled. Thus we end up with a total of 4668 grid nodes around the expected correlation line. 

With the parameter grid defined, we carried out simulations of the Warm Breeze flux observed by IBEX with the parameters from the grid nodes. The simulations were carried out for the entire range of spin angles from the upwind hemisphere, so that we were able to select the spin angle range for the parameter fitting relatively easily. Then, for a selected spin angle range, we compared the simulations with the data using the data covariance matrix and found the best-fit parameters with their covariance matrix as proposed by \citet{swaczyna_etal:15a}. Based on this, we calculated the parameter uncertainties and correlations. In addition to the baseline fit, we performed additional test fits for the full ram hemisphere and for two additional spin angle ranges, one narrower by four data points per orbit, and another one for a spin angle range wider by four data points. This was done to test the robustness of the results. Results of the fitting are listed in Table~\ref{tab:fitResults}.

\subsection{Results and discussion}
\begin{deluxetable}{llllllccccc}
\tabletypesize{\scriptsize} 
\tablecolumns{10}
\tablewidth{0pc}
\tablecaption{Fit results depending on spin angle selection}
\tablehead{
\colhead{Case} & \colhead{$\lambda[\degr]$} & \colhead{$\beta[\degr]$} & \colhead{$v$~[\kms]} & \colhead{$T$~[K]\tablenotemark{b}} & \colhead{$M$} & \colhead{$\xi$} &
\colhead{$N_{\mathrm{dof}}\tablenotemark{c}$} & \colhead{$\chi^2_{\mathrm{min}}$} &
\colhead{$\chi^2_{\mathrm{min}}/N_{\mathrm{dof}}$}} 
\startdata
  $228\degr$--$306\degr$                       & $251.96 \pm 0.65$ & $12.64 \pm 0.34$ & $11.44 \pm 0.57$ & $10\,450 \pm 1\,190$ & $1.90 \pm 0.04$ & $0.062 \pm 0.004$ & 747                & $1443.92$   & $1.933$ \\
	$216\degr$--$318\degr$\tablenotemark{a}      & $251.57 \pm 0.50$ & $11.95 \pm 0.30$ & $11.28 \pm 0.48$ & $9480 \pm 920$       & $1.97 \pm 0.04$ & $0.057 \pm 0.004$ & 963                & $1821.80$   & $1.892$ \\
	$204\degr$--$330\degr$                       & $250.22 \pm 0.45$ & $11.47 \pm 0.30$ & $12.38 \pm 0.48$ & $11\,620 \pm 1\,000$ & $1.95 \pm 0.03$ & $0.065 \pm 0.004$ & 1179               & $2341.95$   & $1.986$ \\
	$180\degr$--$354\degr$                       & $249.39 \pm 0.44$ & $11.40 \pm 0.29$ & $11.98 \pm 0.45$ & $11\,100 \pm 980$    & $1.93 \pm 0.03$ & $0.064 \pm 0.004$ & 1611               & $3312.51$   & $2.056$ \\

\enddata
\tablecomments{The uncertainties obtained from the fits have been scaled up by a factor of $\sqrt{\chi^2_{\mathrm{min}}/N_{\mathrm{dof}}}$ to acknowledge for the values of minimum chi square significantly exceeding the statistically expected values. The uncertainty of the expected value of minimum chi square is equal to $\sqrt{2 N_{\mathrm{dof}}}$.}
\tablenotetext{a}{The baseline spin angle range selection (see text).}
\tablenotetext{b}{Rounded to 10~K.}
\tablenotetext{c}{Number of degrees of freedom in the fit.}
\label{tab:fitResults}
\end{deluxetable}

\begin{figure*}
\centering
\includegraphics[width=1.00 \textwidth]{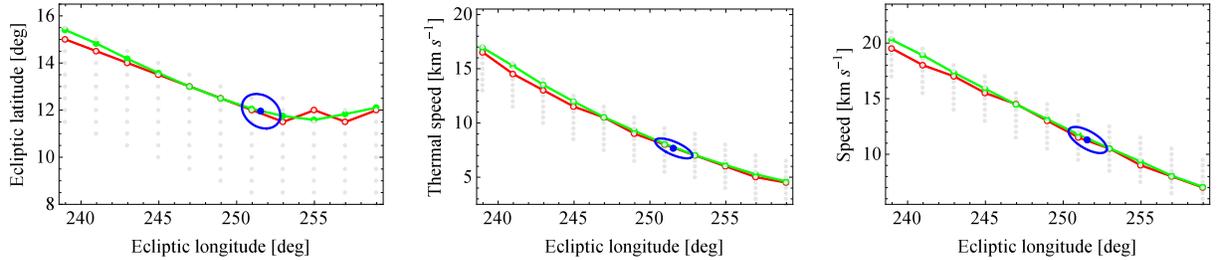}
\caption{Parameter correlation lines projected into 2D subspaces of the 4D parameter space, as a function ecliptic longitude. The gray dots are the simulation grid points. The red line connects the grid points for which the minimum chi square value was obtained for a given longitude. The green line connects the results of inter-grid optimization, i.e., the chi square minima found for the parameter subspaces deployed around the given node in longitude. The blue dots represent the locus of the absolute minimum of chi square listed in the second row of Table~\ref{tab:fitResults} and the blue ellipses are the contours of projections of the $2\sigma$ 4D ellipsoid on the 2D parameter subspaces.}
	\label{fig:corrlines}
\end{figure*}

\noindent Fitting the parameters brought the best-fit solution $\lambda\WB = 251.573\degr$, $\beta\WB = 11.954\degr$, $v\WB = 11.284$~\kms, $v_\mathrm{{T,WB}} = 7.659$~\kms. The temperature is thus $T\WB = 9475$~K and the Mach number of the flow $M\WB = 1.97$. The abundances obtained for individual seasons were $(5.72 \pm 0.29, 6.00 \pm 0.30, 5.79 \pm 0.29, 5.63 \pm 0.30, 5.21 \pm 0.28)\cdot 10^{-2}$. The resulting abundance, calculated as a weighted arithmetic mean value of the abundances obtained for individual seasons, is $\mathbf{\xi\WB = 5.66\cdot10^{-2}}$. The parameters form a ``tube'' in parameter space, similarly as it was found in the case of IBEX observations of the ISN He population. The covariance matrix of the solution is the following:
\begin{equation}
\small
\mathrm{\textbf{Cov}} =\left(
\begin{array}{c|cccc}
& \lambda~\left[\degr\right] & \beta~\left[\degr\right] & v_{\mathrm{T}}~\left[\mathrm{km~s}^{-1}\right]& v~\left[\mathrm{km~s}^{-1}\right]\\ \hline
  \lambda &  0.1301 & -0.01922 & -0.06892 & -0.08179 \\
  \beta & -0.01922 & 0.04622 & 0.01114 & 0.001917  \\
  v_{\mathrm{T}} &  -0.06892 & 0.01114 & 0.07282 & 0.08692 \\
  v & -0.08179 & 0.001917 & 0.08692 & 0.1187  \\
\end{array}
\right).
\label{eq:covMatr}
\end{equation}
This matrix includes the  formal uncertainties resulting from the fitting. These uncertainties are very small, e.g., the uncertainty of the inflow direction is equal to $\sqrt{0.1301}\degr = 0.36\degr$ and the uncertainty of the inflow latitude is $\sim 0.22\degr$. The correlations between the parameters are described by the following correlation matrix:
\begin{equation}
\mathrm{\textbf{Cor}} =\left(
\begin{array}{cccc}
   1     & -0.2479 & -0.7082 & -0.6584 \\
 -0.2479 & 1       &  0.1920 & 0.02588 \\
 -0.7082 & 0.1920  &  1      & 0.9350 \\
 -0.6584 & 0.02588 &  0.9350 & 1 \\
\end{array}
\right),
\label{eq:corrMatr}
\end{equation}
and illustrated in Figure~\ref{fig:corrlines}. The strongest correlation, 0.935, exists between the inflow speed and thermal velocity. The weakest correlation is for the parameter pairs including the inflow latitude (see the second row in the correlation matrix). The inflow longitude is relatively strongly anticorrelated with the thermal velocity, and also with the inflow speed due to the strong correlation of the latter with the inflow speed. A similar pattern of correlations was observed by \citet{bzowski_etal:15a} for ISN He. Projections of the 4D correlation line and of the grid points on 2D subspaces are presented in Figure~\ref{fig:corrlines}. The content and format of this figure is very similar to the format of Figure~5 for the primary ISN He flow in \citet{bzowski_etal:15a}. 

\begin{figure*}
\includegraphics[width=0.49 \textwidth]{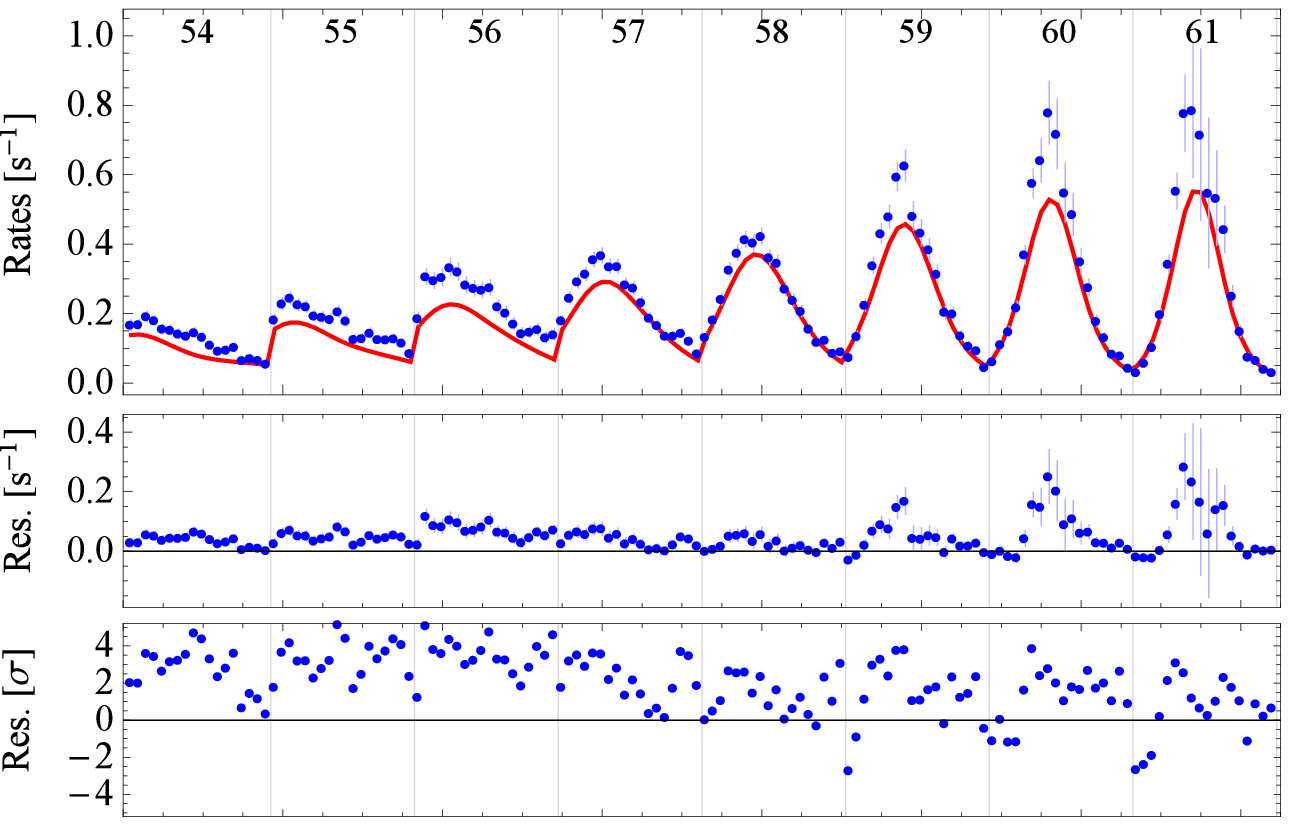} \includegraphics[width=0.49 \textwidth]{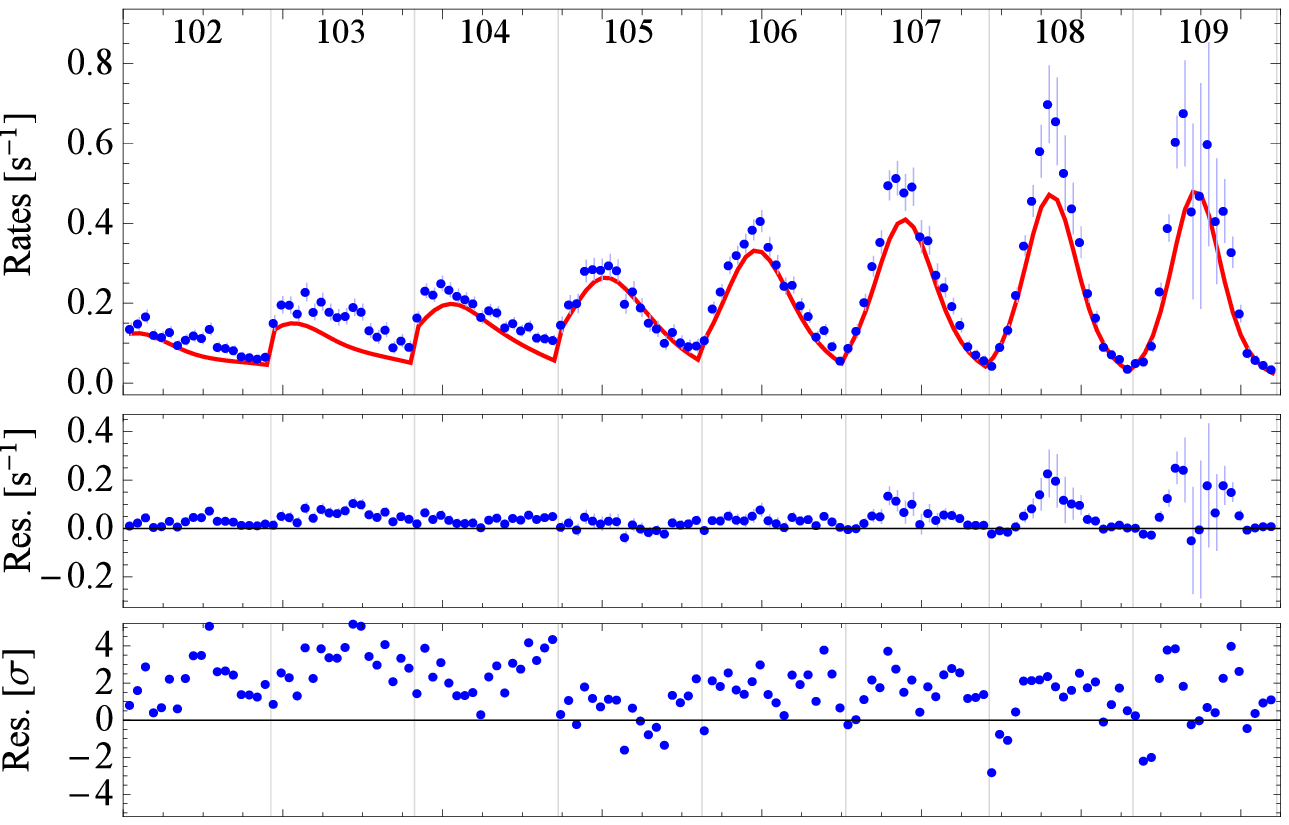}

\includegraphics[width=0.49 \textwidth]{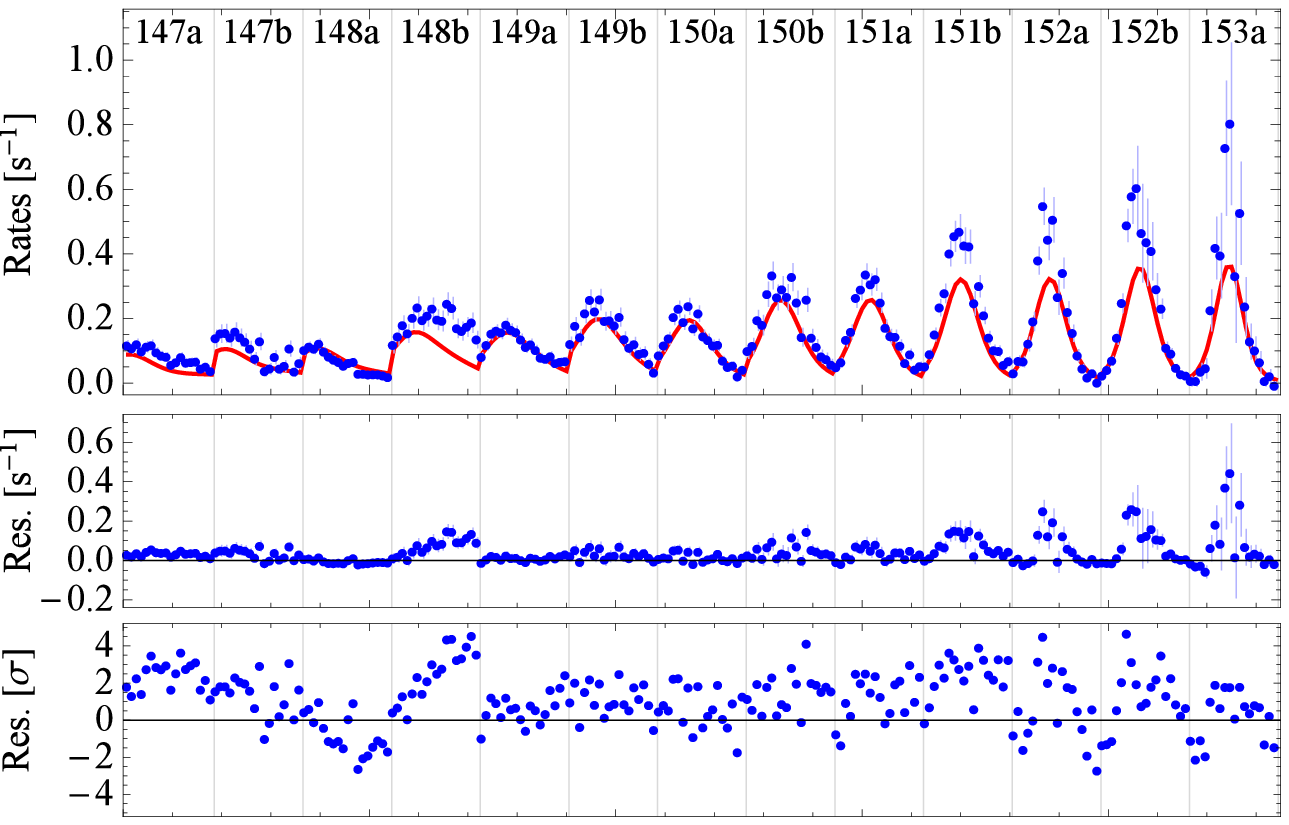} \includegraphics[width=0.49 \textwidth]{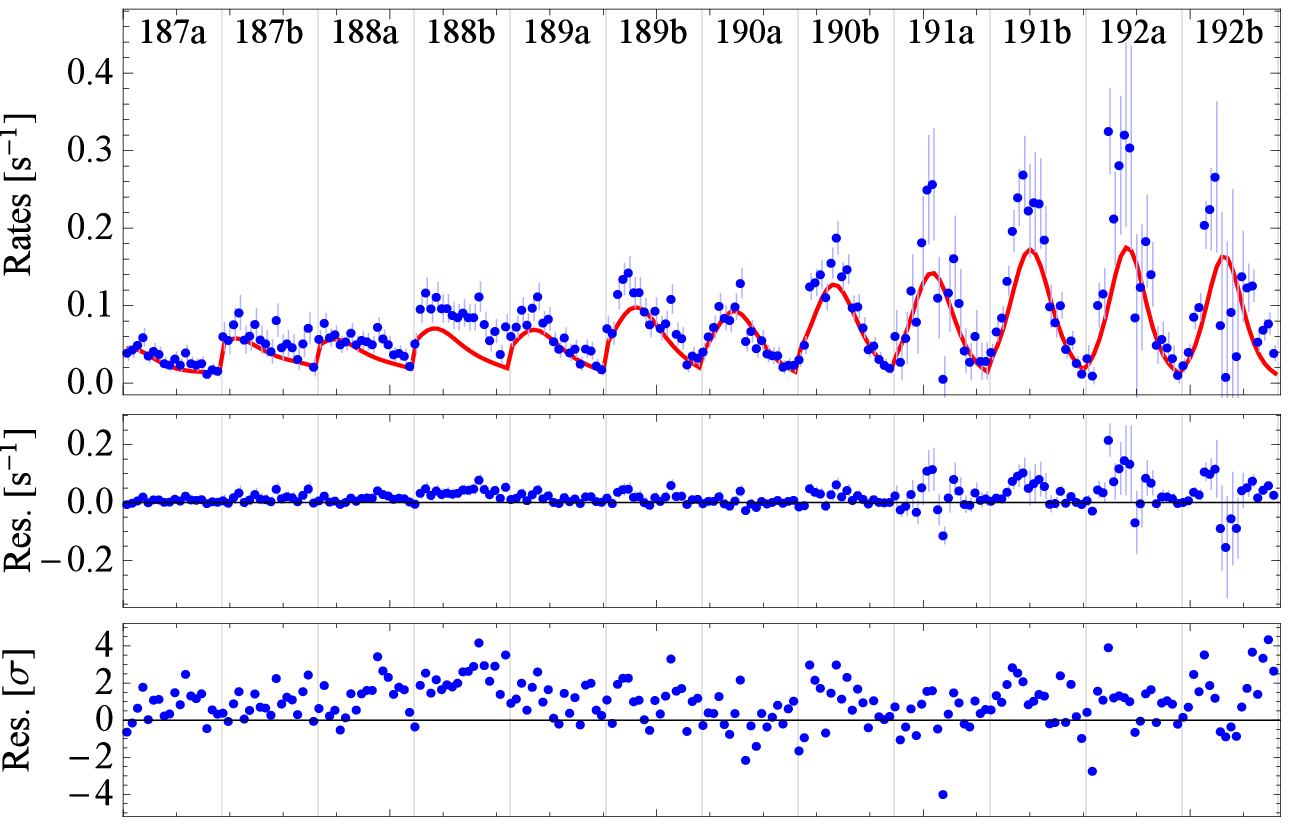}

\includegraphics[width=0.49 \textwidth]{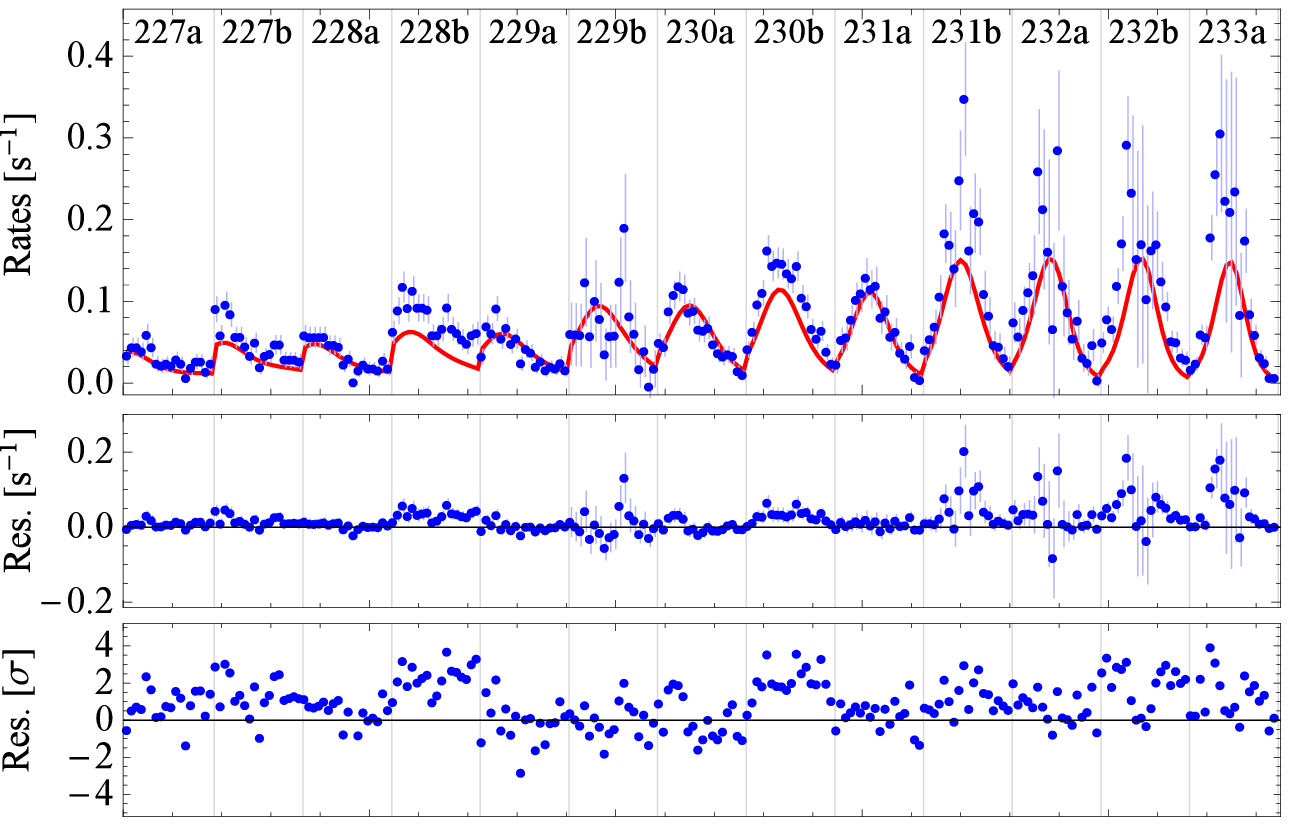}

\caption{Comparison of the data (upper panes, blue dots with error bars) with the best-fit model (red line), and the residuals: absolute (middle pane), and normalized (i.e., the absolute residuals divided by the total uncertainty; lower panes) for all five observation seasons analyzed, from 2010 (upper left) to 2014 (lower left). The vertical bars partition the panels into fragments corresponding to individual orbits, whose numbers are listed at the top of the upper pane of each panel. The horizontal axis is the data point number in the analyzed sample for this observation season. The spin angle range, identical for all of the seasons, is from $216\degr$ to $318\degr$ and one data point corresponds to a 6-degree accumulation bin. }
	\label{fig:residuals}
\end{figure*}

The minimum chi square value obtained from the fitting is equal to $\chi^2_{\mathrm{min}} = 1821.80$ for the number of degrees of freedom $N_{\mathrm{dof}} = 963$, which suggests that the fit quality is unsatisfactory because the minimum chi square obtained is much greater than the statistically expected value, equal to $N_{\mathrm{dof}} \pm \sqrt{2 N_{\mathrm{dof}}} = 963 \pm 43.9$. As discussed by \citet{swaczyna_etal:15a}, a situation where the minimum chi square significantly exceeds the statistically expected value is not unusual in physics and astrophysics. The possible reasons include underestimated data uncertainties, inadequate/incomplete interpretation model, or an additional signal in the data not accounted for in the analysis. Such a situation was also encountered by \citet{bzowski_etal:15a}, who decided to adopt a procedure of scaling up the uncertainties by multiplying them by the square root of reduced chi square, i.e., by $\sqrt{\chi^2_{\mathrm{min}}/N_{\mathrm{dof}}}$. This uncertainty scaling was suggested, among others, by \citet{olive_etal:14a}. In our case, $\chi^2_{\mathrm{min}}/N_{\mathrm{dof}} \simeq 1.9$, so the uncertainties should be multiplied by a factor of $\sim 1.4$. We perform this scaling again here and list the results as the parameter uncertainties in Table~\ref{tab:fitResults}, but we refrain from adopting these uncertainties as the final for our parameters because of the reasons explained later in this section. 

\begin{figure*}
\includegraphics[width=1.0\textwidth]{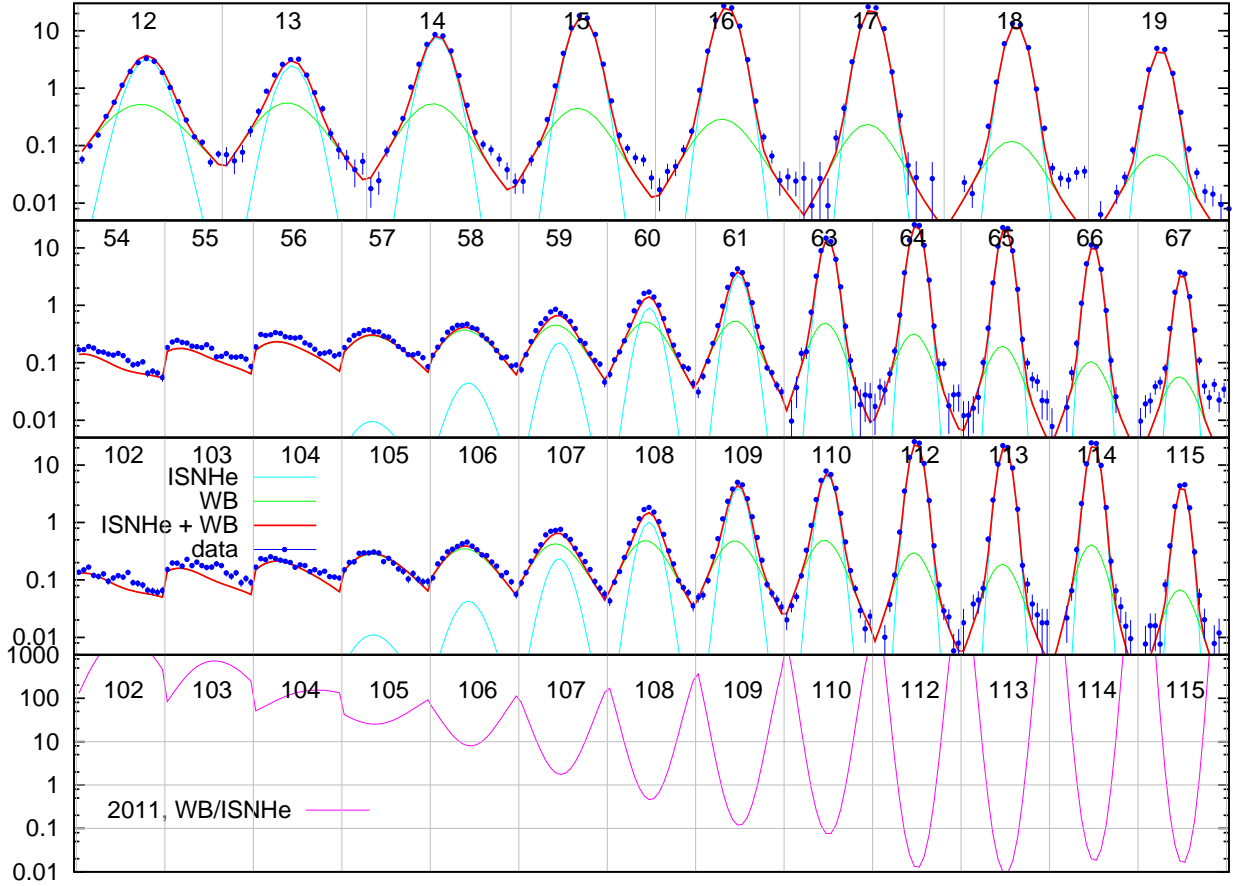}
\caption{Comparison of the data and the full model including the primary ISN He and the Warm Breeze, for the orbits from the 2009---2011 ISN observation seasons used by \citet{bzowski_etal:15a} for the analysis of ISN~He (orbit $\# \geq 14$ for 2009, $\geq 63$ for 2010, and $\geq 110$ for 2011), as well as the available earlier orbits, for the spin angle range $216\degr$---$318\degr$. Note that none of the data points from the 2009 season were used for fitting the Warm Breeze parameters. The spin angle range used by \citet{bzowski_etal:15a} for the ISN He analysis is $252\degr$---$282\degr$, which corresponds to the center six points for orbits 14 through 19 and their equivalents from the later seasons. The data points are arranged in the increasing order of their respective spin angles, the subsets corresponding to individual orbits are partitioned by the vertical bars. The vertical axis is scaled in counts per second. The cyan line represents the model of the primary ISN He, the green line the model of the Warm Breeze, and the red line corresponds to the sum of the latter two components. The blue symbols represent the measured count rates (with the constant background subtracted) and their uncertainties. The lower panel presents the ratio of the flux due to the Warm Breeze to the flux due to ISN He for the 2011 season. For the other seasons, details of this ratio are different, but the general behavior does not change.}
\label{fig:dataVsModel0}
\end{figure*}
\begin{figure*}
\includegraphics[width=1.0\textwidth]{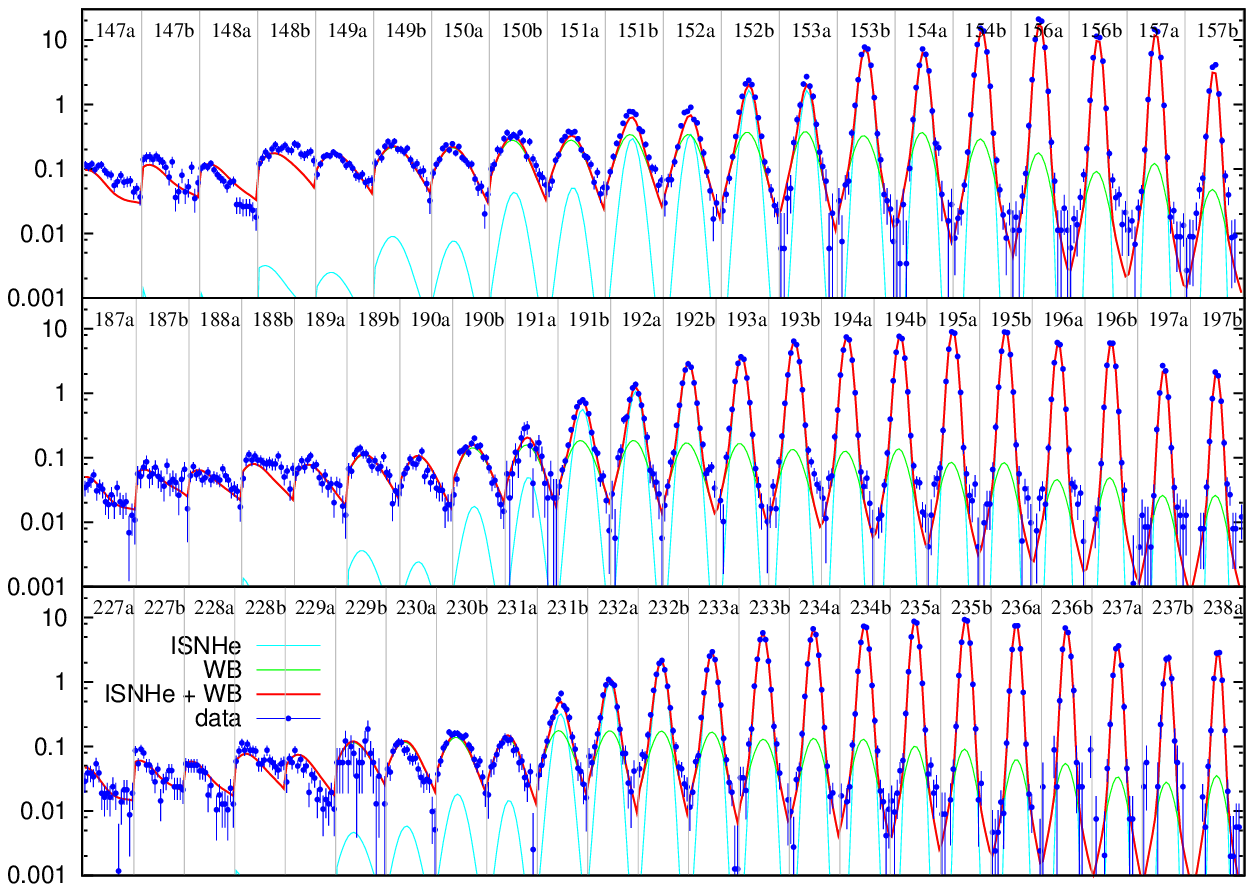}
\caption{Comparison of the data and the full model including the primary ISN He and the Warm Breeze, for the orbits from the 2012---2014 ISN observation seasons used by \citet{bzowski_etal:15a} to fit the ISN He parameters (orbit $\# \geq 153$b, $\geq 193$a, and $\geq 233$b for the three seasons, respectively) and for the orbits used now in the Warm Breeze fitting, for the spin angle range $216\degr$---$318\degr$.  The color and symbol code is identical to that in Figure~\ref{fig:dataVsModel0}. Note the consistently lower levels of the signal in 2013 and 2014 because of the reduction in the PAC voltage introduced after the 2012 ISN season.}
\label{fig:dataVsModel1}
\end{figure*}

Similarly as for the primary ISN population \citep{bzowski_etal:12a, bzowski_etal:15a, mccomas_etal:12b}, the correlation between the parameters results in a very elongated, deep minimum of chi square in parameter space. The blue lines in Figure~\ref{fig:corrlines} illustrate the isocontours for the chi square value equal to $\chi^2_{\rm{min}} + 6.2 \cdot 1.9$, corresponding to the region of a $2\sigma$ uncertainty, scaled by the minimum reduced chi square value to acknowledge the uncertainty scaling descibed previously. All chi square values outside the regions marked by the blue contours in Figure~\ref{fig:corrlines} are larger. 

The direct cause of the high value of minimum chi square can be inferred from inspection of Figure~\ref{fig:residuals}. While the best-fit model generally reproduces the observed signal quite well, two intervals of ecliptic longitudes can be identified where the best-fit solution systematically deviates from the data. The most conspicuous of them begins at $\lambda \simeq 95\degr$ (orbit 59 and the equivalent ones from the subsequent seasons, cf. also Figure~\ref{fig:goodTimes}). Approximately six data points near the center of the spin angle range in those orbits show an excess of the data over the model, which systematically increases with the increasing Earth longitude, even though the normalized residuals do not look that bad in this region because of the large uncertainties of the data there. The remaining data points from these orbits do not show any systematic deviations. While the statistics in this interval is the best during an individual observation season (relatively large count numbers registered), the uncertainty is large because the signal in this region has a large contribution from the primary ISN He population, which is practically absent in the data collected in the earlier portion of the Earth's orbit. An application of the uncertainty system from \citet{swaczyna_etal:15a} results in the relatively large uncertainties in the count rate left for fitting the Warm Breeze after subtraction of the primary ISN He model.

Another region where the residuals are relatively large is the Earth longitude range below $\sim 75\degr$ (orbits 56 and the earlier ones, as well as their equivalents in the following seasons). In this region, the residuals are also predominantly positive, especially in the 2010---2012 seasons, but the statistics of the observed atoms is the lowest in the sample. This excess of the signal over the model can be understood based on analysis by \citet{galli_etal:14a}, who suggested that remnants of the magnetospheric foreground may persist in this region, despite all the filtering procedures applied. An argument in favor of this hypothesis may be a reduction of this phenomenon in the data from 2013 and 2014, i.e., after the reduction of the post-acceleration voltage \citep{mobius_etal:15a}\footnote{Note that the vertical scales in the upper and middle panes of the yearly panels in Figure \ref{fig:residuals} are adjusted to follow the actual amplitude of the signal, which is reduced in 2013 and 2014.}. The magnetospheric contamination is believed to be mostly due to H atoms (the dominant ENAs emitted by the magnetosphere), and their energy seems to mostly be in energy channels 1 and 2. The helium atoms from the Warm Breeze are more energetic. The reduction in the PAC voltage resulted in a decrease in the sensitivity of IBEX-Lo, and the reduction seems to be stronger for the atoms with lower energies. Hence a likely hypothesis to explain the residuals behavior is the disappearance of the magnetospheric foreground in the data because the instrument became less sensitive to this component. 

Another hypothesis, which is complementary to the former one, is that the residual is due to ISN H. \citet{saul_etal:12a} and \citet{schwadron_etal:13a} pointed out that the core of the ISN H contribution to the signal observed in the lowest energy channels of IBEX-Lo is observed in orbit 23 and the corresponding ones during the solar minimum epoch, and that this signal fades during the epoch of high solar activity due to an increased level of the repulsive solar radiation pressure. However, an analysis by \citet{kubiak_etal:13a} suggests that a non-zero flux from ISN H (calculated as a superposition of the primary and secondary populations) is expected throughout the entire ISN observation season (see their Figure~3) and that maximum intensity of this signal for orbits 55 through 58 should be only a little lower than that of the primary ISN He. The ISN H flux for these orbits is expected to be at a level of a few times $10^{-4}$ of the signal expected for ISN He at the seasonal peak intensity, i.e., at a level comparable to the level of the signal from ISN He. The latter one for these orbits, as seen in the lower panel of Figure~\ref{fig:dataVsModel0}, is at a level of 0.01 of the Warm Breeze signal. This makes the hypothetical contribution from the ISN H to the residuals of our present model of the Warm Breeze of a similar order of magnitude to what we actually observe. Furthermore, \citet{kubiak_etal:13a} predict an appreciable reduction of the ISN H flux during the times of high solar activity, and indeed, the residuals that we obtained for the seasons of high solar activity are lower for this portion of the Earth's orbit. This topic certainly deserves further studies, because, on one hand, calculations by \citet{kubiak_etal:13a} were carried out with the assumptions similar to those made by \citet{katushkina_etal:15b}, and on the other hand, \citet{katushkina_etal:15b} showed, using a very sophisticated model of ISN H, that these assumptions lead to a simulated signal very different from the signal actually observed for the orbit where the ISN H flux maximum is expected. To narrow this gap,  these authors had to significantly modify the radiation pressure used in the simulations. The consequences of this modification for the ISN H flux expected on the orbits we discuss now, i.e., 54---57 and the equivalent from the other seasons are unknown. However, further investigation of this aspect is beyond the scope of our paper.

Following the same argument, the excess of the data above the simulation at the large longitude at the end of the data set is likely due to the primary ISN He atoms. In this case, despite the reduction of the PAC voltage and the increased level of solar activity, the excess has not disappeared, so very likely this excess is due to He atoms, not H atoms. A non-perfect reproduction of the ISN He population in our analysis can, in fact, be expected based on the insight provided by \citet{bzowski_etal:15a}. They pointed out that the parameters of the ISN He model they had obtained may be biased due to an imprecise knowledge of the Warm Breeze parameters. These parameters were imprecise, as was suggested by these authors, and as can be seen from our present analysis. On the other hand, statistically speaking, the only evidence for this excess is the visible correlation between the positive absolute residuals for the Earth longitudes larger than $\sim 90\degr$, since the magnitude of the normalized residuals is not significantly larger than in the remaining portion of the data.

The agreement between the data and the model of the neutral He signal is good even for the orbits that were not used in the Warm Breeze fitting. This is illustrated in Figures~\ref{fig:dataVsModel0} and \ref{fig:dataVsModel1}. Even though the ISN He model was fitted only to the six center points for each of the orbits used by \citet{bzowski_etal:15a}, and the Warm Breeze fitting did not use the data from these orbits at all, the agreement between the data and the sum of the ISN He and Warm Breeze populations is evident and most of the small deviations seem to be random. They typically occur at the boundaries of the spin angle range, where a contribution from the local foreground may still be present. Inspection of the lower panel of Figure~\ref{fig:dataVsModel0}, which shows the ratio of the Warm Breeze flux to the ISN He flux, reveals the balance between the two populations in different orbits and different spin angles, and confirms the choice of data by \citet{mobius_etal:15b}, \citet{leonard_etal:15a}, \citet{swaczyna_etal:15a}, \citet{schwadron_etal:15a}, and \citet{bzowski_etal:15a} for their analyses of the primary ISN He flow: the data they used contain relatively little contribution from the Warm Breeze. Simultaneously, it can be seen that a small remnant contribution was still present, as suspected by \citet{bzowski_etal:15a} and \citet{mobius_etal:15b}.

A more fundamental reason of the high value of the chi square minimum may be a weakness of the adopted model for the parent population. We approximate this population with a Maxwell-Boltzmann distribution function with spatially homogenous parameters. But if the Warm Breeze is the secondary population of ISN He, created in the outer heliosheath, then this approximation is certainly not perfect and we expect significant spatial gradients in the flow speed, direction, and temperature of the parent gas. Therefore the parameters we derive in our analysis must be regarded as a kind of mean values, spatially averaged over the source region of the Warm Breeze population. We speculate that such spatial gradients of the parent plasma parameters could be responsible for some systematic departures of the simulated signal from the data and thus for the high value of the minimum chi square found.  

To check the robustness of the solution, we reviewed the fit results for the additional spin angle ranges mentioned earlier (see Table~\ref{tab:fitResults}). The two wider ranges included the portions of the data where the signal is expected to depend on the non-zero energy threshold of the sensitivity of IBEX-Lo, reported by \citet{galli_etal:15a} to be at least $\sim 20$~eV. It is evident that the fit parameters react to the change in the spin angle range adopted for the analysis. The absolute magnitude of the changes is larger than the uncertainties of the fitting, even after scaling them up. This is not surprising, since broadening the range of spin angles includes some data points affected by the uncertain sensitivity of the instrument to low-energy atoms. Relatively, the largest changes are seen in the temperature. Not surprisingly, the reduced chi square minimum, which is a measure of departure of the model from the data per degree of freedom, is larger when we include these additional data and is the lowest for the case we have selected as the baseline, which supports our choice. 

We conclude from this test that the formal uncertainty estimates are too optimistic even after scaling them up to accommodate the high minimum chi square value. Therefore, in addition to the uncertainties resulting from the scaled-up covariance matrix, we also include uncertainties related to the poorly known drop in sensitivity for low energies. These additional uncertainties are estimated as the mean absolute values of the differences between the result of the baseline case and the cases listed in the first and third row in Table~\ref{tab:fitResults}: $\Delta \lambda\WB = 0.9\degr$, $\Delta \beta\WB = 0.6\degr$, $\Delta v\WB = 0.6$~\kms, $\Delta T\WB = 1\,600$~K, $\Delta M\WB = 0.05$, and $\Delta \xi\WB = 0.007$.

Narrowing the uncertainty of the Warm Breeze parameters will be possible after the sensitivity of the IBEX-Lo detector to He atoms with low energies is better understood. This requires carrying out a post-calibration on the spare version of the instrument, which is planned in the near future. With this additional calibration, we will hopefully be able to extend the data range into the spin angle regions affected by the decreasing instrument sensitivity for lower-energy atoms and to use data from two or hopefully three energy channels, which will further improve the statistics. For now, we adopt the uncertainty of the Warm Breeze inflow parameters listed in row 2 in Table~\ref{tab:fitResults}, additionally broadened by $\Delta \lambda\WB$, $\Delta \beta\WB $, $\Delta v\WB$, $\Delta T\WB$, $\Delta M\WB$, and $\Delta \xi\WB$, respectively: $\lambda\WB = (251.57\pm 0.50 \pm 0.9)\degr$, $\beta\WB = (11.95 \pm 0.30 \pm 0.6)\degr$, $v\WB = (11.28\pm 0.48 \pm 0.7)$~\kms, $\xi\WB = (5.7 \pm 0.4 \pm 0.7)\cdot 10^{-2}$. The temperature is $T\WB = (9.48 \pm 0.92 \pm 1.6)\cdot 10^3$~K and the Mach number $M\WB = 1.97 \pm 0.04 \pm 0.05$. The correlations between the uncertainties listed as the first ones are described by Equation~\ref{eq:corrMatr}.

The uncertainty range obtained in the present analysis of five IBEX Warm Breeze observation seasons marginally overlaps with the uncertainty range provided by \citet{kubiak_etal:14a} based on their analysis of the 2010 WB observation season alone. The most likely values obtained now differ in the longitude by $\sim +11\degr$ and in the temperature by $\sim -5500$~K, with the remaining parameters changed very little. The reasons for these differences are most likely 1) the improved data selection we adopted here compared to that by \citet{kubiak_etal:14a}, 2) using by those authors a much less precise version of the model of the ISN He population than currently available (they used the ISN He parameters from \citet{bzowski_etal:12a}), and 3) the fact that the data set used in the present study was larger by a factor of five because now we have data from five observation seasons, not just one. Additionally, whereas \citet{kubiak_etal:14a} used the data from the entire observation season, including the portion where the contribution from the primary ISN He population dominates, here we used solely the portion of the data not used by \citet{bzowski_etal:15a} to fit the primary ISN He parameters.

\section{Implications}

\begin{deluxetable}{llllllccccc}
\tabletypesize{\scriptsize} 
\tablecolumns{4}
\tablewidth{0pc}
\tablecaption{Normal directions to the H and He deflection plane and $\vec{B}-\vec{V}$ plane}
\tablehead{
\colhead{Directions used in fit\tablenotemark{a}} & \colhead{$\lambda[\degr]$} & \colhead{$\beta[\degr]$} & \colhead{$\rho$\tablenotemark{b}}} 
\startdata
  He, WB        & $348.85 \pm 0.83$ & $30.88 \pm 3.89$ & $0.92$  \\
	He, H         & $350.16 \pm 1.50$ & $40.41 \pm 8.02$ & $0.97$  \\
	He, WB, H     & $348.79 \pm 0.84$ & $31.35 \pm 3.85$ & $0.93$  \\
	He, R         & $349.78 \pm 0.60$ & $37.88 \pm 2.61$ & $0.70$  \\
	He, WB, R     & $349.80 \pm 0.57$ & $35.63 \pm 2.06$ & $0.82$  \\
	He, WB, H, R  & $349.70 \pm 0.56$ & $35.72 \pm 2.07$ & $0.85$  \\
\enddata
\tablenotetext{a}{R -- Ribbon, He -- ISN He, H -- ISN H, WB -- Warm Breeze}
\tablenotetext{b}{Correlation coefficient obtained from fit.}
\label{tab:BVPlaneFits}
\end{deluxetable}

\begin{figure}
\centering
\includegraphics[width=0.7 \textwidth]{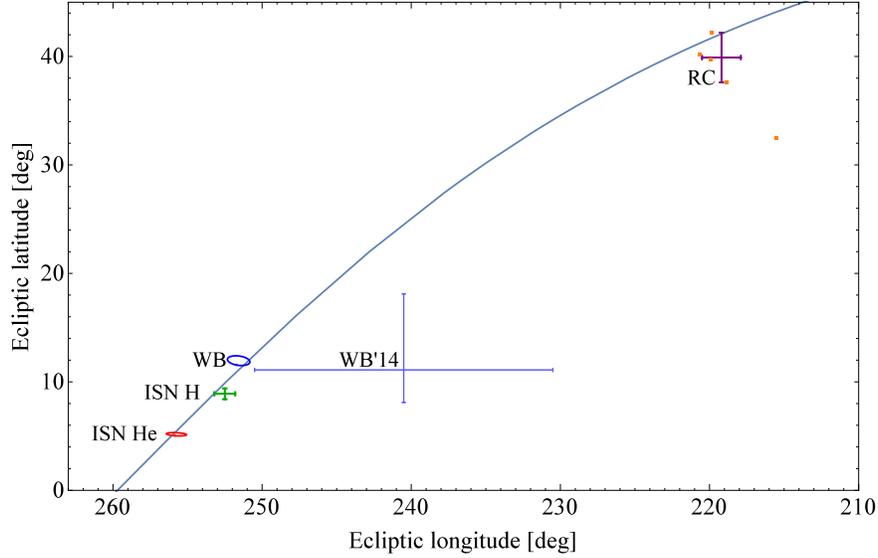}
\caption{Comparison of selected important directions on the sky. WB is the inflow direction of the Warm Breeze from the best-fit model obtained in this paper, with the uncertainty ellipsoid. WB'14 is the Warm Breeze inflow direction obtained by \citet{kubiak_etal:14a}, with the error bars. ISN He denotes the best-fit solution for the ISN He inflow direction obtained by \citet{bzowski_etal:15a} from the analysis of IBEX ISN He observations from 2009---2014. ISN H is the direction of inflow of ISN H with error bars, determined by \citet{lallement_etal:10a} from analysis of SWAN/SOHO observations of the heliospheric backscatter glow; this direction corresponds to the average flow of the primary and secondary ISN H populations. The small orange squares are the directions towards the center of the IBEX Ribbon, determined by \citet{funsten_etal:13a} from observations from IBEX-Hi energy channels 2 through 6 (note they form a monotonic sequence in ecliptic latitude, with the directions for IBEX-Hi energy channels 3 and 4, the closest to the solar wind energy, being the second and third from the top). The purple cross is the average direction for energy channels 2---4, with error bars. The blue line is the great circle fitted to the directions of the Ribbon center, ISN He, ISN H, and the Warm Breeze (see Table~\ref{tab:BVPlaneFits}).}
\label{fig:directions}
\end{figure}

Based on the insight gathered from this study, with all the uncertainties quantified, we can now propose a firm interpretation of the Warm Breeze. This interpretation is based on two indirect pieces of evidence. 

The first of piece of evidence is the magnitude of the deflection of the Warm Breeze inflow direction from the inflow of the primary ISN He inflow. Taking as the basis the ISN He direction obtained recently by \citet{bzowski_etal:15a}, who used a very similar fitting method to the method used here, we obtain the deflection of the Warm Breeze from the primary ISN He equal to $7.9\degr$. Such a deflection, as well as the temperature and inflow speed of the Warm Breeze, are similar to the respective quantities predicted for the secondary ISN He by \citet{kubiak_etal:14a} (see their Figure~11) based on simulations that were carried out using the Moscow Monte Carlo model of the heliosphere \citep{izmodenov_alexashov:15a} with interstellar parameters assumed very close to the parameters currently considered to be the most accurate \citep{bzowski_etal:15a, schwadron_etal:15a, mccomas_etal:15b}.

The other piece of evidence is the observation that the IBEX Ribbon center \citep{funsten_etal:13a} and the directions of inflow of ISN He primary population \citep{bzowski_etal:15a}, ISN H \citep[i.e., a superposition of the primary and secondary populations,][]{lallement_etal:05a, lallement_etal:10a}, and of the Warm Breeze that we have found now are coplanar.

We fitted a great circle on the sky to these four directions, using the uncertainty systems from \citet{bzowski_etal:15a} and the present paper, the Ribbon center and its errors given by \citet{funsten_etal:13a}: $(\lambda_{\mathrm{Ribbon}} = 219.2\degr \pm 1.3\degr, \beta_{\mathrm{Ribbon}} = 39.9\degr\pm 2.3\degr)$ and the ISN H inflow direction and its uncertainty given by \citet{lallement_etal:10a}: $(\lambda_{\mathrm{ISNH}} = 252.5\degr \pm 0.7\degr, \beta_{\mathrm{ISNH}} = 8.9\degr \pm 0.5\degr)$. The fitted great circle is defined by its normal direction in the J2000 heliocentric ecliptic coordinates, equal to $(\lambda = 349\degr \pm 0.6\degr, \beta = 35.7\degr \pm 2.1\degr )$. This circle is plotted in Figure~\ref{fig:directions}. The minimum chi square for this fit is equal to 3.59, while the statistically expected value is equal to $4.0 \pm 2.8$, which implies an excellent fit. As can be seen in Figure~\ref{fig:directions}, the great circle goes through the uncertainty ranges of all four points used in the fit. It is evident from this figure that even if we adopted the WB inflow direction obtained from the fits to the wider ranges of spin angles instead of the one we have actually used, the resulting great circle would still go through the uncertainty ranges of all the points used in the fits, so the co-planarity conclusion would still hold. 

The inflow directions of the ISN He and the Warm Breeze form the so-called Helium Deflection Plane (HeDP), with the normal vector listed in the first row of Table~\ref{tab:BVPlaneFits}. This plane coincides within the uncertainties with the Hydrogen Deflection Plane (HDP), originally suggested by \citet{lallement_etal:05a} and listed in the second row of Table~\ref{tab:BVPlaneFits} for the ISN He direction from \citet{bzowski_etal:15a}, very similar to the derivation by \citet{witte:04}.

As discussed in the {\it Introduction}, this planar alignment of ISN He, ISN H, and the Warm Breeze and of the center of the Ribbon can be naturally explained if the ISMF direction is the direction to the center of the Ribbon and --- simultaneously --- the Warm Breeze is the secondary population of ISN He. Then the Warm Breeze direction is expected to be coplanar with a plane determined by the directions of the local ISMF and the ISN He inflow. To test the robustness of this hypothesis against evidence given by the available data, we calculated the normal directions to the plane fitted to the ISN He inflow directions from \citet{bzowski_etal:15a} and various combinations of ISN H, Ribbon Center, and the Warm Breeze direction obtained here. The results are collected in Table~\ref{tab:BVPlaneFits}. They all agree with each other within their respective uncertainties, which supports the hypothesis that the Warm Breeze is the secondary population of ISN He and the Ribbon center coincides with the ISMF direction.

Adopting this hypothesis, we suggest that the so-called $\vec{B}-\vec{V}$ plane, i.e., the plane including the ISMF vector and the flow vector of interstellar matter is the plane obtained from fitting the directions of the Ribbon center and inflow directions of ISN~He, ISN~H, and the Warm Breeze. The normal vector to this plane is given by the J2000 ecliptic coordinates $\lambda_{\mathrm{BV}} = 349.70\degr \pm 0.56\degr, \beta_{\mathrm{BV}} = 35.72\degr \pm 2.06\degr$, with the correlation coefficient equal to 0.82, as listed in the sixth row in Table~\ref{tab:BVPlaneFits}.

The idea that the deflection of the secondary components of ISN neutrals from the inflow direction of the unperturbed ISN gas is in the plane defined by the velocity vector of the unperturbed ISN gas and the vector of ISMF results from heliospheric models including the interstellar magnetic field and both excluding \citep[e.g.,][]{izmodenov_etal:05a} and including the interplanetary magnetic field \citep{pogorelov_etal:08a}. The inflow direction of ISN H obtained by \citet{lallement_etal:10a} is in fact a superposition of the inflow directions of the primary and secondary populations of ISN H, which are expected to be of comparable densities both in the heliospheric interface and within a few AU from the Sun \citep{katushkina_etal:15b}, where the signal observed by SWAN/SOHO and analyzed by \citet{lallement_etal:05a} is formed. Thus also the ISN H direction is expected to be coplanar with the plane determined by the Ribbon center and the ISN He inflow direction. 

There are essentially two proposed physical mechanisms that create a Ribbon centered on the interstellar magnetic field. The first of these concepts, proposed by \citet{mccomas_etal:09c} and first quantified by \citet{heerkhuisen_etal:10a}, involves the neutral solar wind (i.e., ENAs produced via charge exchange from solar wind protons), which travels out beyond the heliopause and forms a pickup ring after ionization. The mechanism requires that the pickup ring remains stable for long periods (months to years), allowing the pickup ring particles to undergo charge-exchange and generate neutrals. Provided that the neutral particles are produced along the locus where the interstellar magnetic field is roughly perpendicular to the radial direction $(B \cdot r \simeq 0)$, some of these neutrals are directed back toward the Sun and can be observed by IBEX.

Several new lines of evidence also suggest that the Ribbon center is the direction of ISMF. \citet{schwadron_etal:15c} used Voyager 1 observations beyond the heliopause to show that the observed magnetic field steadily rotates, consistently with the “undraping” of the interstellar magnetic field as Voyager 1 moves further out toward the pristine interstellar magnetic field. When the rotation is projected out into the pristine interstellar medium, it is found that the Voyager 1 field direction converges with the center of the IBEX Ribbon. This draping effect was further modeled by \citet{zirnstein_etal:15a, zirnstein_etal:15c} and produces the Ribbon centers close to those observed by IBEX. These authors \citep{zirnstein_etal:15c} found that the direction to the Ribbon center as a function of energy change relatively little and remain in the $\vec{B}-\vec{V}$ plane.

The second line of evidence, indicating consistency between the center of the IBEX Ribbon and ISMF, is found from observations of TeV cosmic rays \citep{schwadron_etal:14a}. In this case, the streaming of cosmic rays determined from TeV cosmic ray anisotropies appears to roughly align with the direction of ISMF determined from the Ribbon center.

The third line of evidence that the IBEX Ribbon center is the direction of ISMF is the consistency of this direction with the interstellar field direction obtained from locally polarized starlight  \citep{frisch_etal:15b}. This implies that the ordering of the interstellar field persists over much larger spatial scales than that of the heliosphere.

\section{Summary and conclusions}

We have analyzed observations of neutral He atoms collected by IBEX-Lo in energy channel 2 during the ISN observation seasons 2010---2014 to estimate the Mach number, temperature, the inflow direction and speed, and the abundance of the Warm Breeze discovered by \citet{kubiak_etal:14a}. We used data collected in the portion of the Earth's orbit that had been excluded from the analysis of the ISN He parameters by \citet{bzowski_etal:15a, leonard_etal:15a, mccomas_etal:15a, mccomas_etal:15b, mobius_etal:15a}, and \citet{schwadron_etal:15a}. We assumed that the observed signal is a superposition of signals due to two Maxwell-Boltzmann populations of neutral He in front of the heliosphere: the primary ISN He population with the parameters known from \citet{bzowski_etal:15a}, and the Warm Breeze population with the parameters we sought to fit. We used a parameter fitting method very similar to the method presented by \citet{swaczyna_etal:15a} and carried out simulations using the Warsaw Test Particle Model, presented by \citet{sokol_etal:15b}, with the time-dependent ionization losses based on the helium ionization history from \citet{sokol_bzowski:14a}. 

We found that the Warm Breeze parameter values obtained directly from the fitting procedure are highly correlated, similarly as it was found by \citet{bzowski_etal:15a} for the ISN He parameters, and that the minimum chi square value significantly exceeds the expected value. We also found that the fit results show some dependence on the data choice because for some spin angles, the observed flux is sensitive to the drop in the sensitivity of the IBEX-Lo instrument to low-energy He atoms, found by \citet{galli_etal:15a} and \citet{sokol_etal:15a}. This additional uncertainty affects mostly the inflow direction and temperature of the Warm Breeze, and is larger than the formal parameter uncertainties obtained from the covariance matrix of the fit. With this additional uncertainty included, the Warm Breeze inflow direction in the J2000 ecliptic coordinates is $(251.57\pm 0.50 \pm 0.9)\degr$, $\beta\WB = (11.95 \pm 0.30 \pm 0.6)\degr$, $v\WB = (11.28\pm 0.48 \pm 0.7)$~\kms. The abundance relative to the primary ISN He is $\xi\WB = (5.7 \pm 0.4 \pm 0.7)\cdot 10^{-2}$, the temperature $T\WB = (9.48 \pm 0.92 \pm 1.6)\cdot 10^3$~K and the Mach number $M\WB = 1.97 \pm 0.04 \pm 0.05$, with the correlations between the uncertainties that are listed as the first ones described by Equation~\ref{eq:corrMatr}; the uncertainties listed as the second entries reflect the uncertainty of the instrument sensitivity. The Warm Breeze parameters obtained in the original derivation by \citet{kubiak_etal:14a} marginally agree with the presently obtained (i.e., the error bars overlap), but the uncertainty obtained now is much smaller.

With the new, more precise direction of the Warm Breeze and with the direction of inflow of ISN He obtained by \citet{bzowski_etal:14a, bzowski_etal:15a, leonard_etal:15a, mccomas_etal:15a, mccomas_etal:15b, schwadron_etal:15a}, and \citet{wood_etal:15a} from IBEX and Ulysses observations we find that these directions are coplanar, within their respective uncertainty ranges. The plane fitted to these four directions is in statistical agreement with a plane containing the directions of inflow of ISN He and the center of the IBEX Ribbon, as well as the plane fitted to the directions of ISN He, ISN H from \citet{lallement_etal:10a}, as well as the plane fitted to the directions of ISN He and WB. Thus the results obtained in this paper for the Warm Breeze, in the papers by \citet{lallement_etal:05a} and \citet{lallement_etal:10a} for ISN H, by \citet{funsten_etal:13a} for the Ribbon center, and by \citet{bzowski_etal:14a, bzowski_etal:15a, leonard_etal:15a, mccomas_etal:15a, mccomas_etal:15b, mobius_etal:15b, schwadron_etal:15a, wood_etal:15a}; and \citet{witte:04} for ISN He are consistent with the hypothesis that the Warm Breeze is the secondary component of ISN He \citep{bzowski_etal:12a, kubiak_etal:14a} and the hypothesis by \citet{mccomas_etal:09c}, \citet{schwadron_etal:09a} and \citet{heerkhuisen_etal:10a} that the direction of the local interstellar magnetic field coincides with the IBEX Ribbon center. This $\vec{B}-\vec{V}$ plane is given by its normal direction in the J2000 ecliptic coordinates $\lambda_{\mathrm{BV}} = 349.70\degr \pm 0.56\degr, \beta_{\mathrm{BV}} = 35.72\degr \pm 2.07\degr$, with the correlation coefficient of 0.85.

\acknowledgments
The authors from SRC PAS acknowledge the support by the Polish National Science Center grant 2012/06/M/ST9/00455. 

\bibliographystyle{apj}
\bibliography{iplbib}{}

\end{document}